\documentclass[aps,pra,showpacs,amssymb,nofootinbib,superscriptaddress,twocolumn,longbibliography,noeprint]{revtex4-2}

\usepackage{comment}
\usepackage[usenames, dvipsnames]{color}
\usepackage{tikz}
\usetikzlibrary{shapes,backgrounds,fit,decorations.pathreplacing}
\usepackage[T1]{fontenc}
\usepackage[utf8]{inputenc}
\usepackage{bbm, bm, amsbsy, amsthm, amsfonts, amsmath, dsfont, graphicx, epsfig, epstopdf, dsfont, multibib, color,relsize,times}
\usepackage{subfigure}
\usepackage{multirow}

\usepackage[colorlinks,breaklinks]{hyperref}
\makeatletter
\newcommand\org@hypertarget{}
\let\org@hypertarget\hypertarget
\renewcommand\hypertarget[2]{%
  \Hy@raisedlink{\org@hypertarget{#1}{}}#2%
  }
\makeatother
\usepackage[figure,table]{hypcap}
\usepackage{MnSymbol}
\usepackage{enumerate}
\usepackage{float}
\hypersetup{
	bookmarksnumbered,
	pdfstartview={FitH},
	citecolor={darkblue},
	linkcolor={darkblue},
	urlcolor={darkblue},
	pdfpagemode={UseOutlines}}
\definecolor{darkgreen}{RGB}{50,190,50}
\definecolor{darkblue}{RGB}{0,0,190}
\definecolor{darkred}{RGB}{238,0,0}
\definecolor{quantum}{RGB}{83,37,127}
\definecolor{quantumlight}{RGB}{169,146,191}
\usepackage{soul}

\usepackage{paralist} %% Only for lists, can be removed in the end

\makeatletter
\def\maketitle{
\@author@finish
\title@column\titleblock@produce
\suppressfloats[t]}
\makeatother

\newcommand{\tr}{\mathrm{Tr}}

\newcommand{\ket}[1]{\ensuremath{\left|\right.\!{#1}\!\left.\right\rangle}}

\newcommand{\bra}[1]{\ensuremath{\left\langle\right.\!{#1}\!\left.\right|}}

\newcommand{\ketbra}[2]{\ensuremath{|{\hspace*{0.75pt}#1\hspace*{0.75pt}}\rangle\!\langle{\hspace*{0.75pt}#2\hspace*{0.75pt}}|}}
\newcommand{\brakket}[3]{\ensuremath{\langle{#1}|{#2}|{#3}\rangle}}

\newcommand{\suptiny}[3]{\ensuremath{^{\hspace{#1 pt}\protect\raisebox{#2 pt}{\tiny{$ #3$}}}}}

\newcommand{\one}{\mathbbm{1}}
\newcommand{\WW}{\mathcal{W}}
\newcommand{\PP}{\mathcal{P}}
\newcommand{\QQ}{\mathcal{Q}}

\newcommand{\expten}[3]{\left\langle #1 \left| \begin{matrix} #2\\ \otimes \\ #3\end{matrix} \right| #1 \right\rangle}

\newcommand{\bigket}[1]{\left|#1\right\rangle}

\renewcommand{\rho}{\varrho}

\begin{document}

\title{Superactivation and Incompressibility of Genuine Multipartite Entanglement}

\author{Lisa T. Weinbrenner}
\affiliation{Naturwissenschaftlich-Technische Fakult\"at, Universit\"at Siegen, Walter-Flex-Stra{\ss}e 3, 57068 Siegen, Germany}

\author{Kl{\'a}ra Baksov{\'a}}
%\email{klara.baksova@tuwien.ac.at}
\affiliation{Atominstitut, Technische Universit{\"a}t Wien, Stadionallee 2, 1020 Vienna, Austria}

\author{Sophia Denker}
\affiliation{Naturwissenschaftlich-Technische Fakult\"at, Universit\"at Siegen, Walter-Flex-Stra{\ss}e 3, 57068 Siegen, Germany}

\author{Simon Morelli}
\affiliation{Atominstitut, Technische Universit{\"a}t Wien, Stadionallee 2, 1020 Vienna, Austria}

\author{Xiao-Dong Yu}
\affiliation{Department of Physics, Shandong University, Jinan 250100, China}

\author{Nicolai Friis}
%\email{nicolai.friis@tuwien.ac.at}
\affiliation{Atominstitut, Technische Universit{\"a}t Wien, Stadionallee 2, 1020 Vienna, Austria}
%\orcid{0000-0003-1950-8640}

\author{Otfried Gühne}
\affiliation{Naturwissenschaftlich-Technische Fakult\"at, Universit\"at Siegen, Walter-Flex-Stra{\ss}e 3, 57068 Siegen, Germany}

\begin{abstract}
Quantum correlations in the form of entanglement, quantum steering 
or Bell nonlocality are resources for various information-processing 
tasks, but their detailed quantification and characterization remain 
complicated. One counter-intuitive effect is the phenomenon of superactivation, 
meaning that two copies of a quantum state may exhibit forms of correlations 
which are absent on the single-copy level. We develop a systematic approach 
towards a full understanding of this phenomenon using the paradigm of genuine 
multipartite entanglement. We introduce systematic methods for studying 
superactivation of entanglement based on symmetries and 
generalized notions of multipartite distillability. With 
this, we present novel criteria for superactivation as well 
as a quantitative theory of it. Finally, we prove the existence of incompressible entanglement, meaning that there 
are quantum states for which superactivated multipartite entanglement cannot be reduced to the single-copy level. 

\end{abstract}

\date{\today}

\maketitle

%%%%%%%%%%%%%%%%%%%%%%
\section{Introduction}
%%%%%%%%%%%%%%%%%%%%%%

The investigation of different quantum resources plays a key role in quantum information 
theory. Much work has been done to characterise, e.g., quantum states with respect to 
their non-locality~\cite{BrunnerCavalcantiPironioScaraniWehner2014}, steerability~\cite{CavalcantiSkrzypczyk2017, UolaCostaNguyendGuehne2020} or entanglement 
\cite{GuehneToth2009, HorodeckiEntanglementReview2009, FriisVitaglianoMalikHuber2019}. 
However, the full characterisation of such resources itself remains a difficult problem 
even in small dimensions. At the same time, experimental realisations allow for manipulating
a steadily increasing number of quantum systems, further complicating the characterisation 
of the associated multi-copy Hilbert spaces.  Another effect adding to the complexity of 
the problem is the superactivation of quantum properties. Superactivation refers to 
phenomena where the combination of objects that individually have none of the desired 
resource results in a joint resourceful object. The resource can thus be activated 
by considering more than one copy. As it was simply put before: \emph{"It is as if $0+0>0$"}~\cite{BrunnerGisin2012}.

Superactivation was first discussed in the context of quantum channels, where the 
combination of two channels with individually vanishing capacity can exhibit a non-zero 
joint channel capacity~\cite{SmithYard2008}. 
A similar effect was also shown in the context of quantum states. There exist local states 
(i.e., whose correlations can be explained by a local hidden-variable model and hence 
satisfy all Bell inequalities) for which two copies become non-local (i.e., can violate 
a Bell inequality)~\cite{Palazuelos2012,SteinbergNguyenKleinmann2025,VillegasAguilarEtAl2024}, 
and there are unsteerable states for which two copies become steerable~\cite{QuintinoBrunnerHuber2016}. 
Recently, also genuine multipartite entanglement (GME) was shown to be activatable~\cite{HuberPlesch2011,YamasakiMorelliMiethlingerBavarescoFriisHuber2022, PalazuelosDeVicente2022, BaksovaLeskovjanovaMistaAgudeloFriis2024}.
Any bi-separable state which is not separable for any fixed partition can be 
activated, meaning that some (potentially large) number of copies of this state 
will be GME. The activation of GME for a noisy Greenberger-Horne-Zeilinger (GHZ) 
state was also investigated experimentally~\cite{ChenEtAl2024}. Besides the 
fundamental interest in this phenomenon, superactivation can be expected to become 
highly relevant for future quantum technologies, as multi-qubit systems are becoming 
easily accessible on many platforms~\cite{FriisMartyEtal2018,CaoEtAl2023,vandeStolpeEtAl2024,Bluvstein-Lukin2024}.

\begin{figure}[t]
    \centering
    \includegraphics[width=0.95\linewidth]{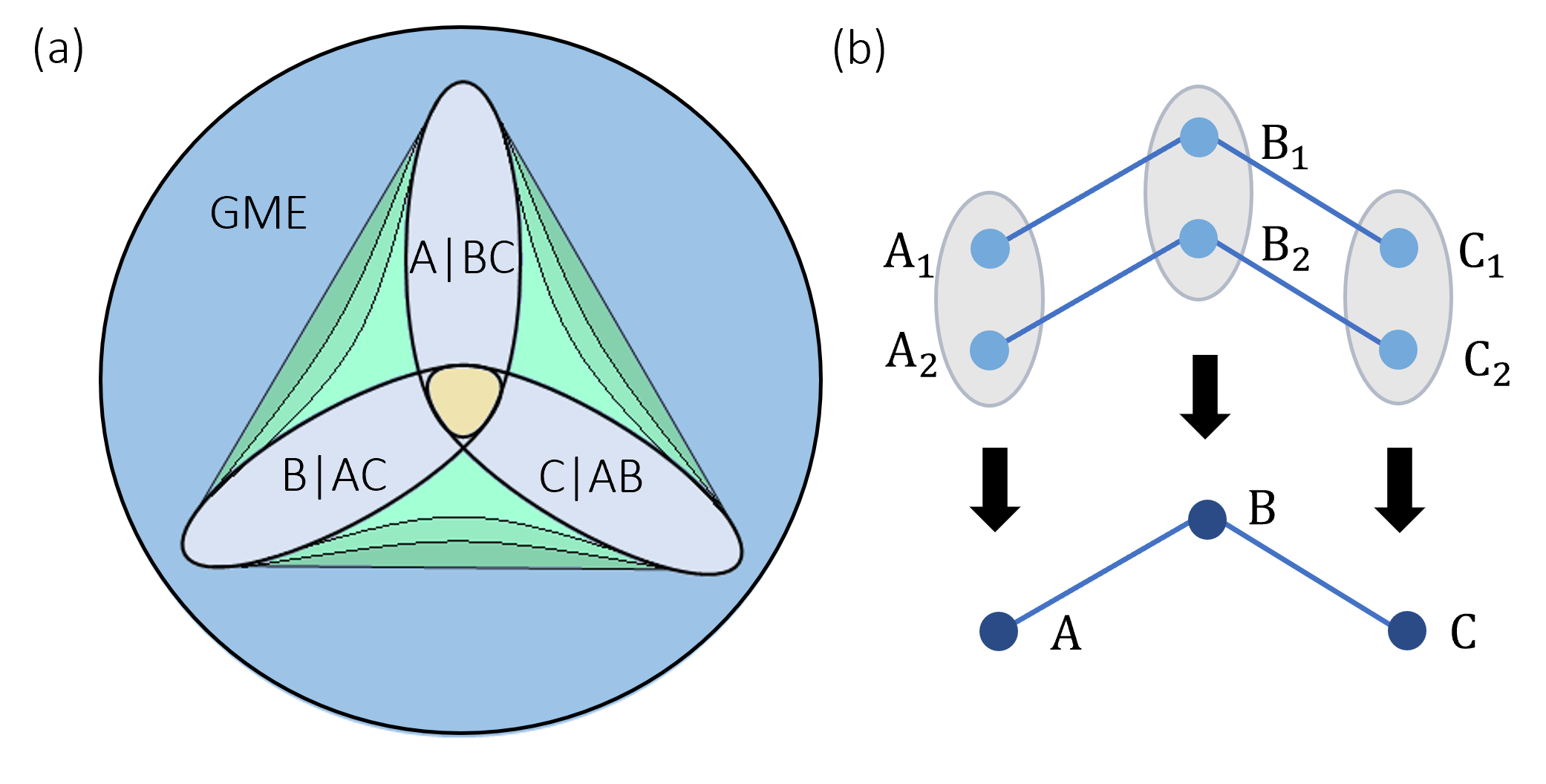}
    \caption{Schematic sketches showing GME activation. (a) Entanglement structure of the tripartite state space. The states which are 
    biseparable, but not separable for a fixed partition (shown in 
    green) can be divided into different 
    subclasses, depending on how many copies of them become GME. 
    {The darker the color the less copies are needed.}
    (b) Two-copy GME activation; two copies of a biseparable state 
    can form a GME state in the two-copy space. This may be shown by
    locally mapping each pair of parties (e.g., $A_1A_2$) to a single, lower-dimensional party ($A$). We prove 
    that there are biseparable states for which two copies are GME, but this entanglement can never be reduced to a single copy again; so it is incompressible.}
    \label{fig:overview}
\end{figure}

The aim of this paper is to develop cornerstones of a general theory 
of superactivation within the paradigm of multi-party quantum 
correlations. To this end we go beyond the simple question of 
existence and towards the characterization and quantification 
of superactivation of GME. Several questions arise naturally 
in this context: First, we ask for the amount of GME that 
can be activated from two or more copies of a biseparable state,
and we present tools to address this question. Second, drawing 
a connection to the distillability problem
\cite{Lewenstein2000Separability}, we ask whether there 
exists activated two-copy GME which can never be reduced to 
the single-copy regime again and present explicit examples
for this phenomenon of incompressible entanglement (ICE).
This can be considered as a multipartite
analogue to phenomena occurring in the search for bound entanglement 
with a negative partial transpose (NPT), a fundamental problem
in quantum information science~\cite{HorodeckiRudnickiZyczkowski2022, 
PankowskiPianiHorodeckiHorodecki2010}. In addition, notions of 
incompressibility are of interest 
in quantum measurement theory~\cite{BluhmRauberWolf2018, UolaKraftDesignolleEtal2021, LoulidiNechita2021}. Finally, we address and answer in the affirmative the question whether superactivation allows it to reach different equivalence classes 
under stochastic local operations and classical communication (SLOCC)~\cite{DuerVidalCirac2000}, which is a widely studied concept in multipartite entanglement theory, see e.g.~\cite{BurchardtQuintaAndre2024,Sauerwein2019,Slowik2020}.

%%%%%%%%%%%%%%%%%%%%%%%%%%%%%%%%%%%%
\section{GME and its activation}
%%%%%%%%%%%%%%%%%%%%%%%%%%%%%%%%%%%%
To start, we recall basic notions of multipartite entanglement
\cite{GuehneToth2009, FriisVitaglianoMalikHuber2019}, 
see also Fig.~\ref{fig:overview}. For simplicity, we  
consider three parties, Alice, Bob and Charlie; the generalization to $N$ parties is straightforward. A 
state $\varrho^{ABC}$ is (i) {\it fully separable} 
if it can be written in the form 
$\varrho^\mathrm{fs} = \sum_i p_i \varrho_i^A 
\otimes \varrho_i^B \otimes \varrho_i^C$, (ii) {\it partition-separable} with 
respect to a given partition, 
e.g., $A|BC$, if it is of the form $\varrho^\mathrm{ps} = \sum_i p_i \varrho_i^A 
\otimes \varrho_i^{BC}$, and finally (iii) {\it biseparable}, if it is a convex 
combination of partition-separable states: $\varrho^\mathrm{bs} = 
p_{\!A}\,\varrho^{A|BC} + p_B\,\varrho^{B|AC} + p_C\,\varrho^{C|AB}$, with $p_{\!A} + p_B + p_C =1$ and $0 \leq p_{\!A}, p_B, p_C \leq 1$.
Lastly, if the state is not biseparable, it is
\textit{genuine 
multipartite entangled} (GME). GME states are frequently considered to be useful, for example in quantum metrology~\cite{Toth2012,HyllusLaskowskiKrischekEtal2012}. Moreover, proving GME has been established as a benchmark for quantum experiments (see~\cite{CaoEtAl2023} for a recent example), because GME implies entanglement between all parties.

Let us assume that the parties share two copies 
of a biseparable state $\varrho_{ABC}=\varrho^{A_1 B_1 C_1} \otimes 
\varrho^{A_2 B_2 C_2}$. Here and in the following, the subscripts $A$, $B$, and $C$ in $\varrho_{ABC}$ indicate that one considers
multiple copies of a state.
If the single-copy state $\varrho^{A_i B_i C_i}$ is partition-separable, it is clear that also the two-copy 
state $\varrho_{ABC}$ must be partition-separable with respect to 
the parties $A=A_1A_2$, $B=B_1B_2$ and $C=C_1C_2$. However, if the 
biseparable state is not partition-separable, the case is not so clear 
since one has to consider the cross terms between the different 
bipartitions. Indeed, it is possible for the two-copy state to become 
GME~\cite{HuberPlesch2011, YamasakiMorelliMiethlingerBavarescoFriisHuber2022}; 
in fact, {\it every} biseparable (but not partition-separable) state 
will become GME for some number of copies $k$~\cite{PalazuelosDeVicente2022}. So, 
every such state is GME activatable, but there are no general bounds 
known on the number of copies needed to activate a certain state 
or the amount of entanglement that can be activated. This leads directly to the following questions: Given a state $\varrho^{ABC}$, 
how many copies $k$ does one need to activate this state? How much entanglement can be activated? And, restricting oneself to the 
two-copy case, how much entanglement can one get by activating a biseparable state? 

%%%%%%%%%%%%%%%%%%%%%%%%%%%%%%%%%%%%%%%%%%%%%%%
\section{Local projections and  distillability}
%%%%%%%%%%%%%%%%%%%%%%%%%%%%%%%%%%%%%%%%%%%%%%%

In the present literature, the main way to detect superactivation 
of GME is the following 
\cite{YamasakiMorelliMiethlingerBavarescoFriisHuber2022}: 
Starting from the two-copy state $\rho_{ABC}$, the parties 
locally apply the \textit{Hadamard map} which is a local projection 
from the two-copy space to the single-copy space. For qubits, the Hadamard map is given by 
$\varrho_{ABC} \rightarrow \mathcal{E}(\varrho_{ABC}) 
= [E_A\otimes E_B\otimes E_C] \varrho_{ABC} 
[E_A^\dagger \otimes E_B^\dagger \otimes E_C^\dagger]$ with 
\begin{align}
\label{eq:Hadamardmap}
E_X &=\, \ketbra{0}{00} + \ketbra{1}{11},
\end{align}
for $X=A,B,C$, where the state $\mathcal{E}(\varrho_{ABC})$
may require renormalization. Applied to each single party, the
action of this map on two quantum states can alternatively
be calculated as
$\mathcal{E}(\varrho\otimes\sigma) \sim \varrho\circ\sigma$ 
where the operation "$\circ$" is the component-wise 
product of two matrices, called Hadamard (or Schur) product.
Since the Hadamard map is a local projection, 
the state $\varrho_{ABC}$ must be entangled if the 
projected state $\mathcal{E}(\varrho_{ABC})$ is entangled, which 
%itself 
can be checked with well-known entanglement criteria.
But are such local projections the only way to characterize
multi-copy quantum correlations? Or are there weak forms of 
correlations which fundamentally escape from this kind of 
detection method? Answering such questions is not only 
essential for experiments~\cite{ChenEtAl2024} but also relevant for interpretational issues~\cite{Wiesniak2024}.

An intriguing first observation is that this scheme is 
directly linked to the well-known problem of entanglement 
distillation~\cite{Lewenstein2000Separability}. Recall that a bipartite state 
$\varrho^{AB}$ is called $k$-copy distillable if there 
exist some local projectors $P$ and $Q$ from the $k$-copy 
space to the qubit space, such that
\begin{align}
\mathcal{D}(\varrho) &=\, (P\otimes Q) \varrho^{\otimes k} 
(P \otimes Q)^\dagger
\label{eq-icedistillation}
\end{align}
is entangled~\cite{Clarisse2006,HorodeckiRudnickiZyczkowski2022}. Clearly, it can be difficult to find appropriate 
projections for a given state $\varrho$, 
in fact, finding such projections for a specific instance
of two copies of a state in dimension $4\times4$ has been flagged as one
of the five central problems in quantum information theory
\cite{HorodeckiRudnickiZyczkowski2022}.

Given the results from the distillability problem it is natural to assume that not every GME-activatable state can be detected by 
using the Hadamard map only. Indeed, as we will 
show below, that one can construct states exhibiting incompressible entanglement,
that is, GME states 
which can never be detected as GME after applying a 
local projection map.  This sparks a series of additional questions: Which states can be 
detected by local projections? How to find the optimal local projection? How can we design useful tests to characterize and quantify superactivation of GME without referring to local maps? Our results are
based on two powerful methods which we explain now.

%%%%%%%%%%%%%%%%%%%%%%%%%%%%%%%%
\section{Core methods}
%%%%%%%%%%%%%%%%%%%%%%%%%%%%%%%%

To start, we reformulate the problem. We demonstrate the idea with two 
copies of a three-qubit state $\varrho$, however, this is directly generalizable to $k$ copies. Assume that we want to detect the entanglement in the two-copy state $\varrho^{\otimes 2}$ by applying local projections 
(like the Hadamard map) and detecting the entanglement in the single-copy space with the witness~$\WW$. This means that we 
search for projections~$F_A$,~$F_B$ and~$F_C$ such that
\begin{align}
\tr\left[ (F_A \otimes F_B\otimes F_C) \varrho^{\otimes 2} (F_A^\dagger \otimes F_B^\dagger \otimes F_C^\dagger) 
\WW \right] < 0 .
\end{align}
Interpreting the $2\times 4$ matrices $F_A$, $F_B$ and $F_C$ as $8$-component vectors $\bra{a}$, $\bra{b}$, and $\bra{c}$, respectively, one finds (similar to~\cite{KrausLewensteinCirac2002, RitzSpeeGühne2019}) that the projections $F_A,F_B,F_C$ exist if and only 
if there is a product vector $\ket{a}\ket{b}\ket{c}$ on $(\mathbbm{C}^8)^{\otimes 3}$
such that 
\begin{align}
\label{eq:alternativformulation}
\brakket{abc}{\varrho^{\otimes 2}\otimes \WW}{abc} < 0 ,
\end{align}
where $\ket{a}$ ($\ket{b}$, $\ket{c}$) acts on system $A$ ($B$, $C$) of $\varrho_{ABC}$.

This formulation can be exploited in two different ways. First, one can use a seesaw algorithm similar to~\cite{GuehneReimpellWerner2007, GerkeVogelSperling2018} to optimize over the product state $\ket{abc}$. 
For this we note that for fixed vectors $\ket{b}$ and $\ket{c}$ the 
optimal $\ket{a}$ can be determined by finding the eigenstate 
corresponding to the minimal eigenvalue of the remaining operator 
on Alice's space. Therefore, one can randomly choose three initial 
states $\ket{a_0}$, $\ket{b_0}$ and $\ket{c_0}$, and then iteratively update each of the three states by fixing the other two. Second, 
one can get a lower bound on the expectation value in Eq.~\eqref{eq:alternativformulation} by relaxing the 
optimisation to run over all states with positive partial transpose (PPT) 
instead of all product states, that means by minimizing the expression
${\tr \left[ \varrho^{\otimes 2}\otimes \WW\ \sigma_{\text{PPT}} \right]}$
over all PPT states $\sigma_{\text{PPT}}$. This can be formulated 
as a semidefinite program (SDP), or in important cases even as a 
simple linear program, see also Appendices~A and~B.
% \supp\nocite{PianiMora2007,Hofmann_2014,Novo_2013,Hein2006,KnillLaflammeMilburn2001,BrowneRudolph2005,ÖzdemirMatsunagaTashimaYamamotoKoashiImoto2011,PanGasparoniUrsinWeihsZeilinger2003,PittmanJacobsFranson2001,VerstreateCirac2004,GabrielHiesmayrHuber2010,LamiHuber2016,HorodeckiMPR1996,HashemiRafsanjaniHuberBroadbentEberly2012,MaChenChenSpenglerGabrielHuber2011,BourennaneEiblKurtsieferGaertnerWeinfurterGühneHyllusBrußLewensteinSanpera2004,GuehneLuGaoPan2007,WeinbrennerPrasannan2024}. 

These methods both make use of local projections to 
detect superactivation. To formulate stronger 
entanglement tests acting directly on the 
multi-copy space, we use two more tools: 
First,  we use 
the method of PPT mixtures~\cite{JungnitschMoroderGuehne2011a}. 
Instead of testing whether the state can be written as a mixture 
of partition-separable states, we test whether the state can be 
written as a mixture of PPT states. This is an SDP, see also Appendix~B. Second, since the dimension of the 
space scales exponentially in the number of copies, we focus 
on GHZ-diagonal, or more generally graph-diagonal states. 
Such states are diagonal in the graph-state basis, which is given 
by the eigenstates of the stabilizers of the graph, see also
Appendix~C.
The symmetry 
of these states allows to also only consider witnesses with the same symmetry, namely graph-diagonal witnesses, reducing the complexity 
of the optimization to a linear program~\cite{Jungnitsch_2011_pra}.

%%%%%%%%%%%%%%%%%%%%%%%%%%%%%%%%%%%%%%
\section{Results}
%%%%%%%%%%%%%%%%%%%%%%%%%%%%%%%%%%%%%%

We consider now three-qubit states, which are diagonal in the GHZ basis, formed by the eight vectors $\{ (\ket{ijk} \pm \ket{\bar{i}\bar{j}\bar{k}})/\sqrt{2}\}_{i,j,k=0,1}$ where $\ket{\bar{b}}$ denotes the bit-flipped value of the qubit $\ket{{b}}$.
In the computational basis, these states are in X-form, having only (nonnegative) entries 
$(\lambda_1,\lambda_2,\lambda_3,\lambda_4,\lambda_4,\lambda_3,\lambda_2,\lambda_1)$ 
on the diagonal and (real) entries 
$(\mu_1,\mu_2,\mu_3,\mu_4,\mu_4,\mu_3,\mu_2,\mu_1)$ 
on the antidiagonal~\cite{GuehneSeevinck2010}. 
Without loss of generality, the diagonal entries can be ordered, 
$\lambda_1\geq \lambda_2 \geq \dots \geq 0$, and from the positivity of the density matrix it follows that $|\mu_i|\leq \lambda_i$. 
We consider here states where $\mu_i=\lambda_i$, using the short-hand notation
$\varrho=\chi(\lambda_1,\lambda_2,\lambda_3,\lambda_4)$, where the normalization of the state is implicitly assumed. 
It is known that GHZ-diagonal states are biseparable if and only if $|\mu_1| \leq \lambda_2 + \lambda_3 + \lambda_4$~\cite{GuehneSeevinck2010}. 
Deciding biseparability of two copies of these states by applying the Hadamard projection as presented in~\cite{YamasakiMorelliMiethlingerBavarescoFriisHuber2022},
one finds that the state $\rho^{\otimes 2}$ is GME if $|\mu_1|^2 > \lambda_2^2 + \lambda_3^2 + \lambda_4^2$.

However, as can be easily seen, the state $\chi(5,4,3,0)^{\otimes 2}$ is not detected by this scheme. 
A careful analysis of this state leads to several insights on the nature of superactivation. 
First, using the PPT mixture approach, we find that the two-copy state $\chi (5,4,3,0)^{\otimes 2}$ is indeed GME, with a white-noise robustness of at least $p_{\rm wnr} \geq 99/136 \approx 0.5928$ meaning that $p [\chi (5,4,3,0)^{\otimes 2}] + ({1-p})\one_{64}/64$ is entangled for all $p > p_{\rm wnr}$.
So, this state tolerates a remarkably high level of noise, showing that the phenomenon of superactivation can be surprisingly strong. 
Moreover, it demonstrates that the approach of the Hadamard map is clearly limited, motivating the tools introduced above. 
In fact, using the formulation of Eq.~\eqref{eq:alternativformulation} and the seesaw algorithm described above, we can find local projections (different from the Hadamard map) such that the projection of $\chi (5,4,3,0)^{\otimes 2}$ can be detected by the GHZ-fidelity witness $\mathcal{W}_{\rm GHZ} = \one/2 - \ketbra{\mathrm{\mathrm{GHZ}_+}}{\mathrm{\mathrm{GHZ}_+}}$, with $\ket{\mathrm{\mathrm{GHZ}_\pm}} = (\ket{000} \pm \ket{111})/\sqrt{2}$, leading to the conclusion that in this case the Hadamard map is not optimal.

\begin{figure}[t!]
    \centering
    \includegraphics[width=0.9\linewidth]{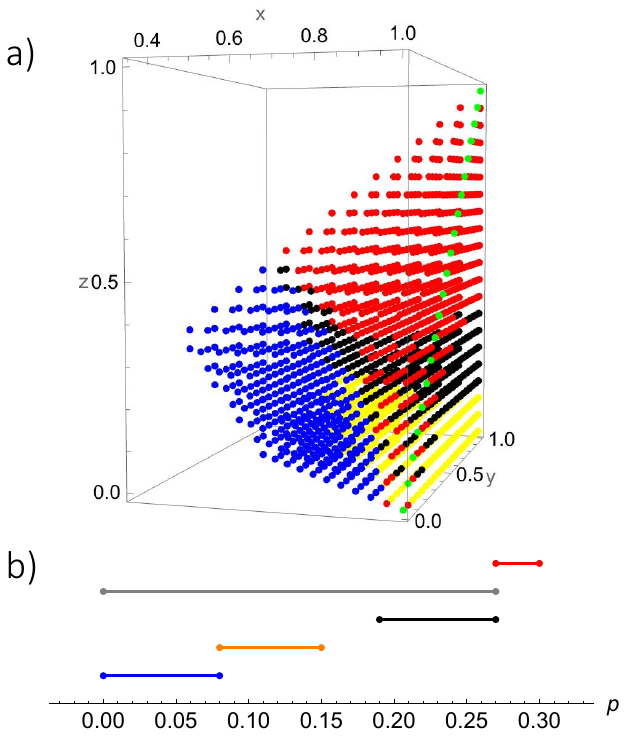}
    \caption{
    (a) Different superactivation properties for 
    noisy GHZ-diagonal states of the type $p\chi(1,x,y,z)+(1-p)\one_{8}/8$ 
    with $p=0.9$ on the two-copy level. States corresponding to green points are 
    separable for a fixed partition and therefore not activatable. 
    The red points are not detected as GME by the approach of PPT 
    mixtures, for all other points superactivation is proven by this approach. 
    The blue points can additionally be detected by the Hadamard map and for the yellow points a different local projection might suffice. 
    For the black points 
    no projections exist which make the states detectable
    by a GHZ-type witness, which is evidence for incompressible 
    entanglement.
    (b) Details on superactivation properties on the two-copy 
    level of the biseparable GHZ-diagonal state $\chi(1,0.65,0.65,0)$ 
    with additional white noise. In the gray (red) area 
    superactivation is (not) found by the PPT mixture; 
    up to 
    $p \approx 0.08$ superactivation can be detected by the 
    Hadamard map (blue) and up to $p\approx0.15$ other local projections 
    were found (orange). From $p \approx 0.019$ on no projections for detection exist. Interestingly, between $p\approx 0.15$ and 
    $p\approx0.19$ no projections could be found, but their existence 
    could also not be ruled out. This may be a signature of PPT
    entanglement, as explained in the text and in Appendix~A.} 
    \label{fig:colobar}
\end{figure}

One may therefore assume that all two-copy entangled GHZ-diagonal states are detectable by a suitable choice of local projections, but this requires a detailed study.
We consider the family $\chi(1,x,y,z)$ of states with varying levels of noise. 
Then the biseparable states form a polytope in the three-dimensional space, subject to the linear constraint $1\leq x + y+ z$.
For these states, we study GME on the two-copy level using PPT mixtures, as well as local projections (or the absence thereof) in combination with the GHZ-fidelity witness. 
The results are displayed in Fig.~\ref{fig:colobar} (and Fig.~2 in Appendix~D). 
The remarkable fact that there are states where two copies are GME, but no local projection makes them detectable on the single-copy level with any GHZ-fidelity witness is strong numerical evidence that these states are indeed ICE on the two-copy level, 
since these witnesses are necessary and sufficient for GME for GHZ-diagonal states. 
For an explicit construction of provably ICE states on two 
copies, which are, however, not GHZ diagonal, we refer to Appendix~G.

\begin{figure}
    \centering
    \includegraphics[width=0.9\linewidth]{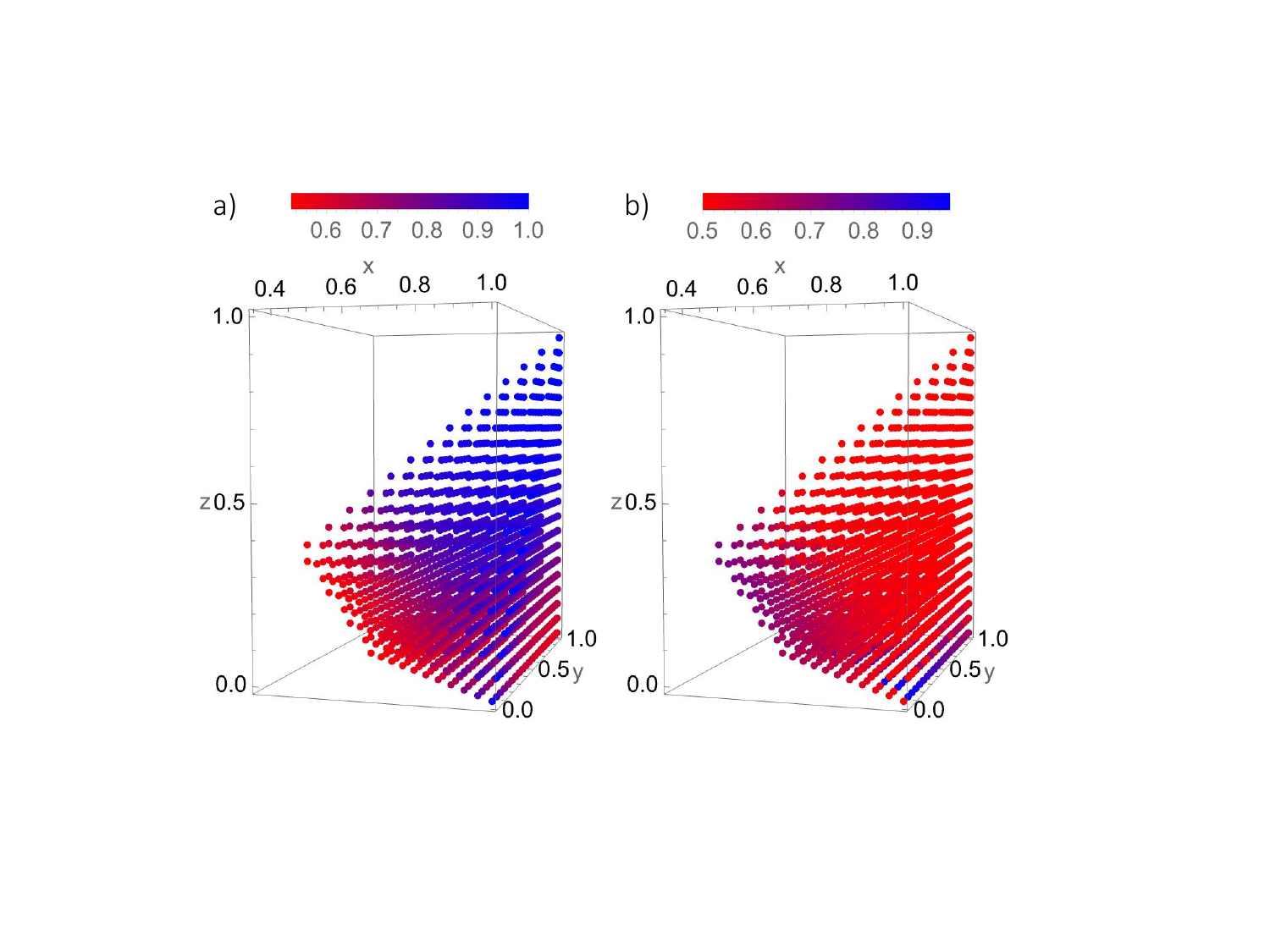}
    \caption{Quantitative results for superactivation of GHZ-diagonal states of the type $\chi(1,x,y,z)$. (a) The white-noise robustness 
    on the two-copy space estimated from the results of the PPT mixture approach
    (b) The three-qubit GHZ fidelity after applying an optimized projection onto the single-copy space. 
    See text for further details.
    \label{fig:3dwhitenoise}
    }
\end{figure}

The example of the state $\chi(5,4,3,0)$ raises the question {\it how much} GME can be generated in two copies of a biseparable state.
Using the results from the PPT mixture approach, we obtain lower bounds on the white-noise robustness, which are displayed in Fig.~\ref{fig:3dwhitenoise}.
The most robust GHZ-diagonal state with a two-copy white-noise robustness of at least $p_{\rm wnr} \geq 0.5294$ is the state $\chi(1,1/3,1/3,1/3)$, which is the most symmetric biseparable GHZ-diagonal state with respect to the diagonal entries $\lambda_i$. 
Note that this state also violates the criterion obtained by the Hadamard map above maximally.

Another potential quantifier of the superactivation phenomenon 
is the GHZ fidelity that can be reached after projecting two copies 
to the single-copy Hilbert space. 
The state $\chi(1,1/3,1/3,1/3)$ considered above reaches a GHZ fidelity 
of $3/4 = 0.75$. 
Interestingly, however, much higher fidelities can be reached from
biseparable states close to partition-separable states by using
local maps {\it different} from the Hadamard map. For instance, a
fidelity of $F=0.97561$ can be reached by taking two copies of 
the state $\chi(1,1,0.05,0)$ and this state is very close to the 
partition-separable state $\chi(1,1,0,0)$. This proves two things:
First, since a GHZ fidelity larger than $0.75$ implies GHZ-type 
entanglement~\cite{AcinBrussSanperaLewenstein2001}, it shows that the phenomenon of
superactivation can result in states belonging to the GHZ class where
entanglement measures like the three-tangle do not vanish~\cite{CoffmanKunduWootters2000}. Second, the resulting operations may pave the way to
novel purification protocols of GHZ states, shedding light on the question
how many copies are needed to go from one SLOCC entanglement class to the other~\cite{DuerBriegel2007, YuChitambarGuoDuan2010}.
Details are discussed in Appendix~D.

\begin{figure}[t]
    \centering
    \includegraphics[width=0.95\columnwidth]{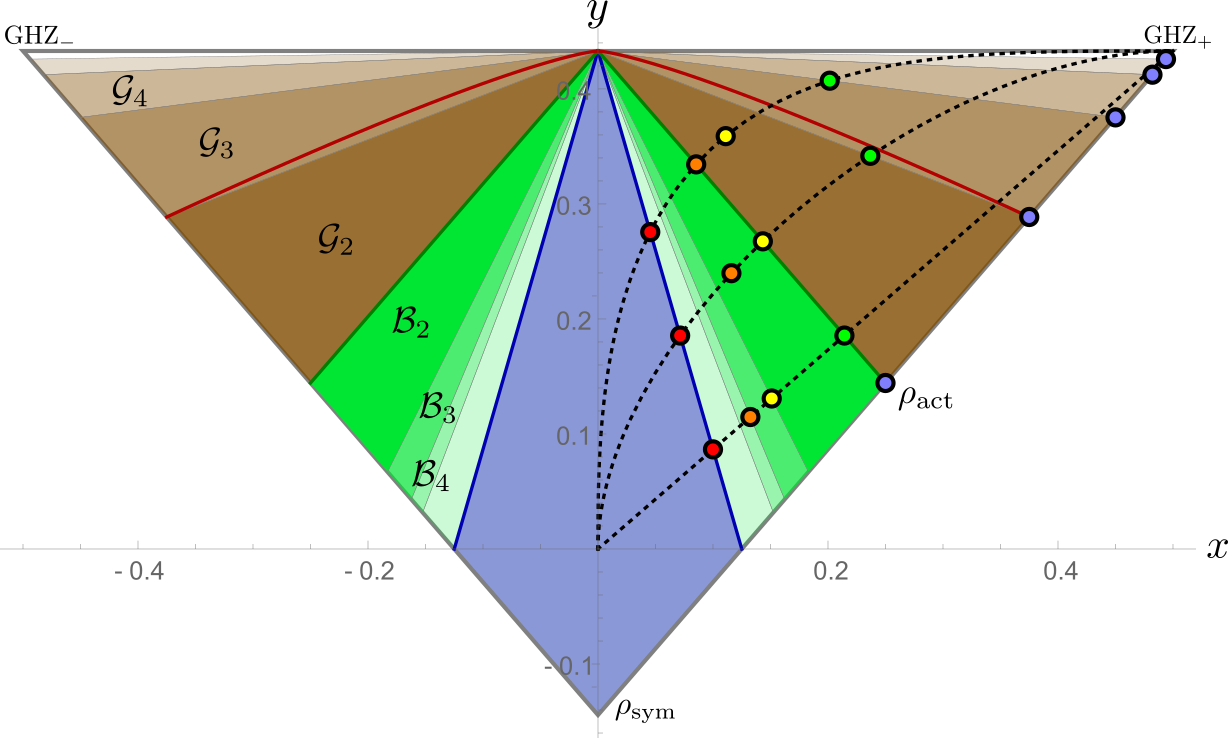}
    \caption{
    {GME activation of three-qubit GHZ-symmetric states}. 
    The plot illustrates the family of three-qubit GHZ-symmetric states, parametrized by 
    $x[\rho]=(\sum_{m=\pm}m\bra{\mathrm{GHZ}_m}\rho\ket{\mathrm{GHZ}_m})/2$ and 
    $y[\rho]=(-{1}/{4}+\sum_{m=\pm}\bra{\mathrm{GHZ}_m}\rho\ket{\mathrm{GHZ}_m})/\sqrt{3}$. 
    The only non-activatable states in this family are fully separable (blue), while biseparable states (green) are not partition-separable and thus activatable. 
    The subregions $\mathcal{B}_2$, $\mathcal{B}_3$ and $\mathcal{B}_4$ of $2$-, $3$- and $4$-copy activatable states are indicated in different shades. 
    Brown regions correspond to GME states, with states below the red line belonging to the W class, and states above belonging to the GHZ class. 
    The shade of brown indicates the number of copies of biseparable states required to reach the respective area via the Hadamard projection. 
    Under the $k$-copy Hadamard map, the region $\mathcal{B}_2$ is mapped to $\mathcal{G}_{k}$. 
    Under the two-copy Hadamard map, the regions $\mathcal{B}_{3}$ and $\mathcal{B}_{4}$ are mapped to $\mathcal{B}_{2}$. 
    The black dotted lines correspond to the isotropic GHZ states and the projection of two and three copies of them from bottom to top. 
    The dots mark special states and their multi-copy Hadamard projections: 
    the first fully separable isotropic state (red dots), the first isotropic state that needs four (orange dots), three (yellow dots), and two copies (green dots) to be activated, respectively. 
    The blue dots on the boundary mark multiple copies of the biseparable state $\rho_\mathrm{act}=\bigl(\ketbra{\mathrm{GHZ}_+}{\mathrm{GHZ}_+} +\rho_{\mathrm{sym}}\bigr)/2$.
    }
    \label{fig:GHZsymmetric_all}
\end{figure}

%%%%%%%%%%%%%%%%%%
\section{GME Activation \& SLOCC Equivalence}
%%%%%%%%%%%%%%%%%%

Having established that superactivation can in 
principle lead to GHZ-type entanglement, let us study
this phenomeon further in a setting where everything can
be treated analytically. 
More specifically, we focus on GHZ-symmetric states 
of three qubits, together with the Hadamard map, 
which maps GHZ-symmetric states to GHZ-symmetric states 
and for which numerical data suggests to be optimal in 
this scenario.

Three-qubit GHZ-symmetric states \cite{DuerCirac2001, EltschkaSiewert_2012a,EltschkaSiewert_2012b,EltschkaSiewert_2012c} form a subset of GHZ-diagonal states and can be described by convex combinations of the states $\ket{\mathrm{GHZ}_+}$, $\ket{\mathrm{GHZ}_-}$, and 
$\rho_{\mathrm{sym}}=(\one_8-\ketbra{000}{000}-\ketbra{111}{111})/6$.
For this family of states, the parameter regions of full separability, biseparability, W-type entanglement and GHZ-type entanglement have been fully
described~\cite{EltschkaSiewert_2012a,EltschkaSiewert_2012b,EltschkaSiewert_2012c}, see Fig.~\ref{fig:GHZsymmetric_all} and its caption.
Concerning the SLOCC classification, one can clearly see that the states that can be reached with two copies ($\mathcal{G}_2$) are a subset of the W class. 
Note also that the blue dots in Fig.~\ref{fig:GHZsymmetric_all} show a sequence of $k$ copies of a state on the GME boundary that approaches the GHZ state. 
A detailed discussion is given in Appendix~E.

%%%%%%%%%%%%%%%%%%%%%%%%%%%%%%%%%%%%%%%%%%%%%%%%%%%
\section{Experimental detection of superactivation}\label{sec:experimentaldetection}
%%%%%%%%%%%%%%%%%%%%%%%%%%%%%%%%%%%%%%%%%%%%%%%%%%%

So far, we have encountered several interesting superactivation phenomena. 
The question remains how superactivation can be observed experimentally.

As a first method, one can use nonlinear entanglement witnesses to prove the superactivatability of a quantum state. 
These can be constructed from a witness certifying GME on the two-copy space.
This witness may be derived from the PPT-mixture approach, other separability criteria or by reversing the projection approach in Eq.~(\ref{eq:Hadamardmap}): 
One can look for a witness on the single-copy state, and then lift it to the two-copy space by inverting the projection map. 
Then, to measure the witness in practice, one needs to decompose it into locally measurable observables.
For instance, any observable on two copies of three qubits can be decomposed into sixfold tensor products of Pauli matrices, corresponding to local observables on each qubit. 
These then result in quadratic but locally measurable expressions on the single-copy space.
In Appendices~E and~F we give explicit examples of such nonlinear criteria and also discuss in detail the statistical analysis
of a potential experimental setup.
Also, using non-Hadamard projections gives further directly measurable criteria for the single-copy state (see Eq.~(D12)) in Appendix~D). 
In Appendix~C it is shown how such ideas can be generalized to arbitrary graph states.

As a second, experimentally more demanding method, one may directly prepare two copies of a state, apply some local projections, and see whether the resulting single-copy state is entangled, as was demonstrated in \cite{ChenEtAl2024}.
It would be especially interesting to perform the distillation steps leading to high-fidelity GHZ entanglement (Eq.~(D12) in Appendix~D) in practice.

%%%%%%%%%%%%%%%%%%%%
\section{Conclusion}
%%%%%%%%%%%%%%%%%%%%

The superactivation of quantum resources is, at first sight, a counter-intuitive phenomenon. 
Yet, it is well established for many types of quantum correlations and it is likely to become relevant if large and multi-copy quantum systems become technologically available.
Our study revealed several interesting phenomena in this context. 
First, we quantified the phenomenon, showing for instance that sometimes already two copies of a biseparable state can lead to GHZ-type entanglement of high purity. 
Second, we provided evidence for incompressible entanglement in GHZ-diagonal states and also proved the existence of the phenomenon of ICE in a different family of
states.
This may be seen as an analogue of bound entanglement, where high-dimensional entanglement cannot be concentrated to two-level systems. 
Finally, we provided concepts to observe superactivation experimentally.

Our methods and results may stimulate further research in several directions. 
An operational characterization of the phenomenon of incompressible entanglement would be highly desirable.
In fact, this may lead to a fresh perspective on the problem of NPT bound entanglement, one of the fundamental problems in quantum information theory that has evaded being solved for more than two decades. 
Independent of that, the study of potential projection procedures from two copies to the single-copy level may be fruitful to improve distillation protocols. 
Finally, a detailed experimental demonstration of superactivation phenomena could lead to new directions in quantum information processing.

%%%%%%%%%%%%%%%%%%%%%%%%%%%
\section*{Acknowledgements}
%%%%%%%%%%%%%%%%%%%%%%%%%%%

We thank Dagmar Bru{\ss}, Carlos de Gois, Kiara Hansenne, 
Hermann Kampermann, Robin Krebs, Ties-Albrecht Ohst, and Jens Siewert 
for discussions.  
This work has been supported by the Deutsche Forschungsgemeinschaft 
(DFG, German Research Foundation, project numbers 447948357 and 440958198),
the Sino-German Center for Research Promotion (Project M-0294), 
the legacy of J.S. Bach (BWV 248),
the German Ministry of Education and Research 
(Project QuKuK, BMBF Grant No. 16KIS1618K), 
the Austrian Federal Ministry of Education, Science, and Research via the Austrian Research Promotion Agency (FFG) through the flagship project HPQC 
(FO999897481), the project FO999914030 (MUSIQ), and the project 
HDcode funded by the European Union - NextGenerationEU. 
L.T.W. and S.D. acknowledge support by the House of Young Talents of the University of Siegen.
K.B. and N.F. acknowledge support from the Austrian Science Fund (FWF) through the project P 36478-N funded by the European Union - NextGenerationEU.
X.D.Y. acknowledges support by the National Natural Science Foundation of China (Grants No. 12205170 and No. 12174224) and 
the Shandong Provincial Natural Science Foundation of China (Grant No. ZR2022QA084).

\newpage

%%%%%%%%%%%%%%%%%%%%%%%%%%%%%%%%%%%%%%%%%%%%%%%%%%%%%%%%%%%%%%%%%%%%%

\onecolumngrid
\appendix

%%%%%%%%%%%%%%%%%%%%%%%%%%%%%%%%%%%%%%%%%%%%%
\section{Numerical procedures for finding optimized projections}
%%%%%%%%%%%%%%%%%%%%%%%%%%%%%%%%%%%%%%%%%%%%%
\label{sec-app-finding-projections}

We argued in the main text that the search for local projections which enable 
the detection of GME in the single-copy space by the linear entanglement witness 
$\WW$ can be reformulated as an optimization problem over product vectors. 
Namely, one can find such projections if and only if there exist some product 
vector $\ket{a}\ket{b}\ket{c}$, such that 
\begin{align}\label{eq:AppCvalue}
C(a,b,c) &:=\, \expten{abc}{\varrho^{\otimes 2}}{\WW} < 0 .
\end{align}

One possible way of finding such product vectors is given by a seesaw 
algorithm, which is similar to the algorithm used in the context of the 
geometric measure of entanglement~\cite{GuehneReimpellWerner2007, 
GerkeVogelSperling2018}. In this algorithm the three vectors $\ket{a}$, 
$\ket{b}$ and $\ket{c}$ are updated iteratively to reach a minimum 
$C_{\text{min}}$. It is important to note that for fixed vectors 
$\ket{b}$ and $\ket{c}$, the vector $\ket{a}$ minimizing the 
expression $C(a,b,c)$ is given by the eigenstate corresponding 
to the minimal eigenvalue of the remaining operator on Alice's space,
\begin{align}
\expten{b c}{\varrho^{\otimes 2}}{\WW} \bigket{a} &=\, \lambda \bigket{a}.
\end{align}
The same argument holds also for $\ket{b}$ and $\ket{c}$. Therefore, one can 
choose arbitrary vectors $\ket{a_0}$, $\ket{b_0}$ and $\ket{c_0}$, 
and then iteratively update these vectors by fixing two of them and solving 
the eigenvalue problem for the third vector. Since the updated vectors are always 
chosen as the eigenvector to the smallest eigenvalue, the value $C(a_i,b_i,c_i)$ 
never increases over each iteration. Naturally, this scheme does not always lead 
to a global minimum, so it may be repeated with different starting vectors. 
As an example, running the algorithm for two copies of a three-qubit 
GHZ-diagonal state $\chi(1, x, y, z)$ by repeating the scheme with $10$ starting
points and $30$ iterations takes about three seconds (using Mathematica on a standard 
laptop).

It should be clear from the description above that this algorithm is not 
proven to find a global minimum. However, the problem of finding suitable 
product states can be relaxed to PPT states which are PPT with respect to 
each partition. Denoting the operator in Eq.~\eqref{eq:AppCvalue} by 
$\mathcal{X} = \varrho^{\otimes 2} \otimes \WW$, we note that 
\begin{align}
\min_{\ket{abc}} \ \brakket{abc}{\mathcal{X}}{abc} 
    &=\, \min_{\rho \text{ fully separable}} \tr[\rho \mathcal{X} ]
    \geq \min_{\rho \text{ PPT}} \tr[\rho \mathcal{X} ]
\end{align}
since all separable states are PPT with respect to every partition. As 
explained in Appendix~\ref{sec-pptmix}, the problem of optimizing over PPT states and 
obtaining the value $\min_{\rho \text{ PPT}} \tr[\rho \mathcal{X} ]$ can 
be solved by a SDP. If  $\mathcal{X}$ has additional symmetries (like being 
three-copy GHZ diagonal) it even reduces to a linear program. If this value 
should be positive, it is clear from the above argument that no projections 
fulfilling Eq.~\eqref{eq:AppCvalue} exist. 

Finally, let us comment on the examples of states $\varrho$, for which
no local projection could be numerically found, but also their existence could not
be excluded by the SDP relaxation (see the discussion in the caption
of Fig.~2 in the main text). Technically, this 
indicates that the operator $\mathcal{X} = \varrho^{\otimes 2} \otimes \WW_{\rm GHZ}$
acting on the tripartite $8 \times 8 \times 8$ space of three copies 
has a positive expectation value on all product states, but a 
negative one on some states being PPT for all bipartitions. This 
just means that $\mathcal{X}$ is a non-decomposable witness, 
detecting some PPT states. Indeed, witnesses of a similar structure 
(but mainly on two-copy spaces) have been discussed before~\cite{PianiMora2007}, 
so the existence of states $\varrho$ with the mentioned properties may have 
been expected.

%%%%%%%%%%%%%%%%%%%%%%%%%%%%%%%%%%%%%%%%%%%%%
\section{The method of PPT mixtures}
\label{sec-pptmix}
%%%%%%%%%%%%%%%%%%%%%%%%%%%%%%%%%%%%%%%%%%%%%

 In this Appendix, we give a short introduction into the method of PPT mixtures~\cite{JungnitschMoroderGuehne2011a}. We also discuss how this 
can be efficiently evaluated in practice for GHZ-diagonal states (and more generally, graph-diagonal states) as mentioned in the main text. We mainly formulate the theory for three parties, generalizations
to more parties are straightforward.  

As mentioned in the main text, a quantum state of three parties is biseparable, if it can be written as
\begin{align}
\varrho^\mathrm{bs} &=\, p_{\!A}\,\varrho^{A|BC} + p_B\,\varrho^{B|AC} 
+ p_C\,\varrho^{C|AB}
\end{align} 
with $p_{\!A} + p_B + p_C =1$ and $0 \leq p_{\!A}, p_B, p_C \leq 1$ and
$\varrho^{A|BC}$ etc.~being biseparable for the respective 
bipartition. The idea of the PPT-mixture approach is to relax 
the condition of biseparability, by taking PPT states instead, 
\begin{align}
\varrho^\mathrm{pptmix} &=\, p_{\!A}\,\varrho^{A|BC}_\mathrm{ppt} + p_B\,\varrho^{B|AC}_\mathrm{ppt} 
+ p_C\,\varrho^{C|AB}_\mathrm{ppt},
\end{align} 
where $\varrho^{A|BC}_\mathrm{ppt}$ etc.~are now PPT for the respective
partitions. Clearly, any biseparable state is a PPT mixture, but not necessarily the other way round. So, if one can prove that a state is not a PPT mixture, it must be GME. In fact, the PPT mixure approach has been shown to provide a necessary and sufficient criterion for many families
of states, such as GHZ-diagonal states or permutationally invariant
three qubit states~\cite{Hofmann_2014, Novo_2013}.  

The interesting point is that the question whether a state is a PPT mixture or not 
can directly be decided via a semidefinite program (SDP). This is most conveniently formulated in the dual picture of entanglement witnesses.
An observable of the form
\begin{align}
\WW   = &  \PP_{A} +\QQ_A^{T_A} = \PP_B +\QQ_B^{T_B} = \PP_C +\QQ_C^{T_C}
\nonumber
\\[1mm]
\mathrm{  with } \quad & \PP_X \geq 0 \mbox{ and } \QQ_X \geq 0 
\mbox{ for all } X \in \{ A,B,C \}
\label{eq-pptmixwitness}
\end{align}
is clearly positive on all PPT mixtures. In fact, for a given 
state $\varrho$ one can minimize $\tr(\varrho \WW)$ among all 
witnesses obeying the constraints from Eq.~(\ref{eq-pptmixwitness})
via an SDP, and this is negative if and only if the state $\varrho$
is not a PPT mixture. Here, one has to normalize $\WW$ in some form.
Fixing $\tr(\WW)=1$ allows one to directly estimate the white-noise robustness
of the GME~\cite{JungnitschMoroderGuehne2011a}, while a normalization of
the type $0 \leq \PP_X $ and $0 \leq \QQ_X \leq \one$ leads to a quantitative interpretation in terms of the mixed convex roof of the bipartite negativity~\cite{Hofmann_2014}.  

For our approach in the main text it is essential that
the approach simplifies further if the states are GHZ
diagonal. Such states obey the symmetry
\begin{equation}
\varrho = g_i \varrho g_i \mbox{ with } g_i \in 
\{Z_1 Z_2 \one_3, X_1 X_2 X_3, \one_1 Z_2 Z_3 \},
\end{equation}
where $X_i$ denotes the Pauli matrix $\sigma_x$ acting on qubit $i$ 
etc., and symmetrizing an arbitrary state to this form
can be implemented by randomly applying the local unitaries $g_i$~\cite{DuerCirac2001, GuehneSeevinck2010, Jungnitsch_2011_pra}. Consequently, these
symmetries can also be applied to the witness $\WW$ 
and the $\PP_X$ and $\QQ_X$ from Eq.~(\ref{eq-pptmixwitness}). If all these operators are diagonal,
checking their positivity  reduces to a linear program (LP).  

In practice, we implemented the SDP using the solver {\tt Mosek} and the LP using the solver {\tt Gurobi}. 
To give an example of the performance, checking the 
PPT-mixture approach for two copies of a GHZ-diagonal state (that is, running the SDP on a $4 \times 4 \times 4$ system) takes approximately 10 minutes, while the linear program requires approximately 3 seconds. All programs are available upon reasonable request.

%%%%%%%%%%%%%%%%%%%%%%%%%%%%%%%%%%%%%%%%%%%%%
\section{Graph-diagonal states}
%%%%%%%%%%%%%%%%%%%%%%%%%%%%%%%%%%%%%%%%%%%%%
\label{sec-app-graphstates}

In this Appendix, we discuss the generalization of some
of the concepts from GHZ-diagonal states to more general
graph-diagonal states. First, we give a brief introduction to
the formalism of graph states. Then, we ask how the Hadamard map
can be meaningfully extended to this more general class of states. We explain the graphical interpretation of 
the Hadamard map in the computational basis as well as 
the analoguous map in the eigenbasis of $\sigma_x$. Finally, 
we demonstrate that this can be used to extend the approach 
from 
Ref.~\cite{YamasakiMorelliMiethlingerBavarescoFriisHuber2022} 
from GHZ states to arbitrary two-colorable graph states. 
This approach made use of the fact that the Hadamard map 
projects $k$ copies of the GHZ state to a GHZ state again and 
therefore the application of a 
GHZ state specific entanglement criterion is possible.

%%%%%%%%%%%%%%%%%%%%%%%%%%%%%%%%%%%%%%%%%%%%%
\subsection{Graph-state formalism}
%%%%%%%%%%%%%%%%%%%%%%%%%%%%%%%%%%%%%%%%%%%%%
First, we give a short introduction to graph states, the reader familiar with them may skip this section. For a graph with $N$ nodes $i\in\{1,\dots,N\}$ 
which are connected by the edges $(i,j)\in E$ we can define the according 
graph state by 
\begin{align}
\ket{G} = \prod_{i,j \text{ s.t. } (i,j)\in E} C_{i,j} \ket{+}^{\otimes N},
\label{eq-graph-state-def-1}
\end{align}
where $C_{i,j} = CZ$ denotes the controlled $Z$ gate on the qubits $i$ and $j$. It reads 
\begin{align}
CZ = \begin{pmatrix}
\ 1 & \ 0 & \ 0 & \phantom{-}0 \\
\ 0 & \ 1 & \ 0 & \phantom{-}0 \\
\ 0 & \ 0 & \ 1 & \phantom{-}0 \\
\ 0 & \ 0 & \ 0 & -1
\end{pmatrix}.
\end{align}
For example, the easiest graph of two nodes, connected by an edge, 
corresponds to the graph state
\begin{align}
C_{1,2} \ket{++} &=\, C_{1,2} \frac{1}{4}( \ket{00} + \ket{01} + \ket{10} + \ket{11}) \,=\, \frac{1}{4} (\ket{00} + \ket{01} + \ket{10} - \ket{11}) = 
\frac{1}{\sqrt{2}}(\ket{0+} + \ket{1-}),
\end{align}
being equivalent to a Bell state.   

A different view on these states can be obtained by using the stabilizer
formalism. For that, one starts with a graph and defines for any node
$i$ the operator
\begin{equation}
g_i = X_i \bigotimes_{j \in \mathcal{N}(i)} Z_j    
\end{equation}
where $X_i$ denotes the Pauli matrix $\sigma_x$ acting on qubit $i$ 
etc., and  $\mathcal{N}(i)$ is the neigbourhood of node $i$ in the 
graph. This procedure gives $N$ commuting observables $g_i$ and the 
graph state can be defined as a common eigenstate of all of these, 
\begin{equation}
\ket{G} = g_i \ket{G}
\label{eq-graph-state-def-2}
\end{equation}
One can directly calculate that these two definitions are equivalent.   

Indeed, one can generalize this to form an entire graph-state basis. 
In Eq.~(\ref{eq-graph-state-def-1}) one may replace $\ket{+}$ by 
$\ket{-}$ on some qubits. There are $2^N$ possibilities of doing 
that, leading to $2^N$ orthogonal (but locally equivalent) states,
the graph-state basis. Equivalently, one can choose in 
Eq.~(\ref{eq-graph-state-def-2}) different eigenvalues $-1$
instead of $+1$ for some of the indices $i$, leading to exactly the 
same basis.  

Given the graph-state basis, one can consider density matrices or 
observables diagonal in this basis \cite{Hein2006}. This is interesting for 
several reasons. First, one can make a quantum state diagonal in this 
basis by applying local operations. Consequently, entanglement
of the graph-diagonal state implies entanglement of the original state.
Second, checking positivity of a graph-diagonal operator is simple, as
one only has to look at the entries on the diagonal of the matrix. 
Third, the partial transposition of a graph-diagonal operator is
graph diagonal, too. The second and third points imply that the method
of PPT mixtures reduces to a linear program and can be solved efficiently. 

\subsection{Generalized Hadamard maps}

 First, one can directly see that not only the Bell state, but also
the GHZ state $\ket{\mathrm{GHZ}} = (\ket{000} + \ket{111})/\sqrt{2}$ falls into
the category of graph states. This state is the common eigenstate of 
the set of stabilizers $Z_1 Z_2 \one_3$, $X_1 X_2 X_3$ and 
$\one_1 Z_2 Z_3$. Applying a Hadamard transformation on qubits 1 and 3
one obtains the operators $X_1 Z_2 \one_3$, $Z_1 X_2 Z_3$ and 
$\one_1 Z_2 X_3$. But this corresponds to the stabilizing operators of
the linear graph with three nodes, where 1 is connected with 2, and
2 is connected with 3.   

The main question then is: What does the Hadamard map do on the graphical description of the GHZ state? How can this be generalized to more 
(arbitrary) graph states? Recall that the action of the Hadamard map on one party in the computational basis is given by the effects
\begin{align}
E^z = \ketbra{0}{00} + \ketbra{1}{11},
\end{align}
see Appendix \ref{app-sec:GabrielWitness} for a more formal definition in terms of density matrices.
For the GHZ state, this clearly maps the stabilizers to stabilizers again, 
if it is applied to two copies of a state. If one considers
the standard formulation of graph states, however, this is not so clear. 
It is then natural to consider the corresponding transformation in the $X$ basis, 
\begin{align}
    E^x = \ketbra{+}{++} + \ketbra{-}{--},
\end{align}
where $\ket{+}$ and $\ket{-}$ denote the eigenstates of $X$. We will explain here how these two operations are interpreted in a graphical way.

%%%%%%%%%%%%%%%%%%%%%%%%%%%%%%%%%%%%%%%%%%%%%%%%%%%%%%%%%%%%%%%%%%%%%%%%%%%
\subsection{Action of the operations $E^z$ and $E^x$ in terms of graphs}
%%%%%%%%%%%%%%%%%%%%%%%%%%%%%%%%%%%%%%%%%%%%%%%%%%%%%%%%%%%%%%%%%%%%%%%%%%%

Now we take a closer look at the two operations $E^z$ and $E^x$ and give an interpretation in graphical terms. We note that this is connected to the operation of graph-state fusion~\cite{KnillLaflammeMilburn2001, BrowneRudolph2005, ÖzdemirMatsunagaTashimaYamamotoKoashiImoto2011}, more precisely we will see that the action of the Hadamard map $E^z$ corresponds to the so-called parity check operation~\cite{PanGasparoniUrsinWeihsZeilinger2003, PittmanJacobsFranson2001,VerstreateCirac2004}. 

We start with the Hadamard operation in Z-basis $E^z$ and assume without loss of generality that it is applied 
to qubits $1$ and $2$. Note, that here we are only interested in the case where node $1$ and $2$ are not connected.
Obviously $E^z$ maps two different qubits to only one resulting qubit, 
so the two nodes $1$ and $2$ will be mapped to a single node $\{1,2\}$. 
There are two different types of nodes connected to $1$ and $2$: There are 
nodes which are connected to both $1$ and $2$, so they are common neighbours of $1$ and $2$ (denoted by the set $V_{1,2}$), and nodes which are only connected to one of these two. These sets are denoted by $V_1$ and $V_2$. Considering the corresponding graph state $\ket{G}$ we are interested in the graph state $E_{1,2}^z \ket{G} = \ket{\tilde{G}}$ obtained after the action of the Hadamard map on qubits $1$ and $2$.   

In the graph state $\ket{G}$ the qubits which are not $1$ and $2$ and 
are not in the sets $V_{1,2}$, $V_1$ or $V_2$ are not effected by 
$E^z$, so we can restrict ourselves without loss of 
generality to graph states of the form
\begin{align}
\ket{G} = 
\prod_{i \text{ s.t. } (1,i)\in E} C_{1,i} \prod_{j \text{ s.t. } (2,j)\in E} C_{2,j} \ket{+}^{\otimes N} 
= \prod_{i \in V_{1}} C_{1,i} \prod_{j \in V_{2}} C_{2,j} \prod_{k \in V_{1,2}} C_{1,k}C_{2,k} \ket{+}^{\otimes N}.
\end{align}
Applying $E_{1,2}^z$ to this graph state essentially picks 
out all terms where we have $\ket{00}_{1,2}$ or $\ket{11}_{1,2}$ on 
the first two qubits. Now we consider the two cases explained above: 
If node $i$ is connected to both $1$ and $2$, then it gets no phase 
from $\ket{00}_{1,2}$ and "two phases" from $\ket{11}_{1,2}$, which 
cancel. So the new node $\{1,2\}$ has no effect on node $i$ and 
therefore there is no edge between these nodes. However, if $i$ 
is only connected to one of $1$ or $2$, then it gets a phase from 
$\ket{11}_{1,2}$. In this case the new node $\{1,2\}$ is connected 
by an edge to $i$. The new graph state $\ket{\tilde{G}}$ is therefore
\begin{align}
\ket{\tilde{G}} = E_{1,2}^z \ket{G} = \prod_{i \in V_{1}} C_{\{1,2\}, i} \prod_{j \in V_{2}} C_{\{1,2\}, j} \ket{+}^{\otimes N-1} .
\end{align}
All in all this yields the following rule:  

\noindent
\textbf{Graphical rule for $E^z$:}
Applying $E_{i,j}^z$ to two unconnected nodes $i$ and $j$ maps these two nodes to one new node $\tilde{i}$. This new node is disconnected from nodes which are neighbours of both $i$ and $j$, and connected to all nodes, which lie in the neighbourhood of either $i$ or $j$.    

Now we look at the action of the map in X-basis $E^x$ on the two 
qubits $1$ and $2$. Obviously this operation also fuses the two nodes 1 and 2 to a single 
node $\{1,2\}$. In fact, one can directly see that it maps 
$\ket{00}_{1,2} \mapsto \ket{0}_{\{1,2\}}$,
$\ket{01}_{1,2} \mapsto \ket{1}_{\{1,2\}}$,
$\ket{10}_{1,2} \mapsto \ket{1}_{\{1,2\}}$, and
$\ket{11}_{1,2} \mapsto \ket{0}_{\{1,2\}}$.
Again, we consider only the case where nodes $1$ and 
$2$ are not connected. Further we assume that these two nodes only have common 
neighbours $V_{1,2}$, meaning that every node connected to either 
$1$ or $2$ is also connected to $2$ or $1$, respectively.
As above, we consider a graph state 
$\ket{G}$ which contains without loss of generality only the nodes $1$, 
$2$ and $i \in V_{1,2}$, and determine the graph state 
$E_{1,2}^x \ket{G} = \ket{\tilde{G}}$ obtained after the 
action of $E_{1,2}^x$ on qubits $1$ and $2$. 
The graph state $\ket{G}$ is then of the form
\begin{align}
    \ket{G} = \prod_{k \in V_{1,2}} C_{1,k}C_{2,k} \ket{+}^{\otimes N} 
    = \prod_{k \in V_{1,2}} C_{k,1}C_{k,2} \ket{++}_{1,2}\otimes \ket{+}^{\otimes N-2}.
\end{align}
Now the terms $\ket{00}_{1,2}$ and $\ket{11}_{1,2}$ have clearly the 
same sign (as well as the terms $\ket{01}_{1,2}$ and $\ket{10}_{1,2}$).
This implies that the resulting state after the map from above can 
directly be interpreted as a graph state. In addition, the signs of
$\ket{0}_{\{1,2\}}$ and $\ket{1}_{\{1,2\}}$ are different. This leads 
to
\begin{align}
    \ket{\tilde{G}} &=\, E_{1,2}^x \ket{G} = E_{1,2}^x \prod_{k \in V_{1,2}} C_{k,1}C_{k,2} \ket{++}_{1,2}\otimes \ket{+}^{\otimes N-2} \,=\, \prod_{k \in V_{1,2}} C_{k,\{1,2\}} \ket{+}_{\{1,2\}}\otimes \ket{+}^{\otimes N-2} 
\end{align}
This argument yields then the following rule:  

 \textbf{Graphical rule for $E^x$}
If two unconnected nodes $i$ and $j$ share the same neighbourhood, 
then applying $E_{i,j}^x$ maps these two nodes to one new 
node $\tilde{i}$. This new node is connected to all common neighbours of $i$ and $j$.

%%%%%%%%%%%%%%%%%%%%%%%%%%%%%%%%%%%%%%%%%%%%%
\subsection{Consequences for two-colorable graph states}
%%%%%%%%%%%%%%%%%%%%%%%%%%%%%%%%%%%%%%%%%%%%%

\begin{figure}[t]
    \centering
    \includegraphics[width=0.75\linewidth]{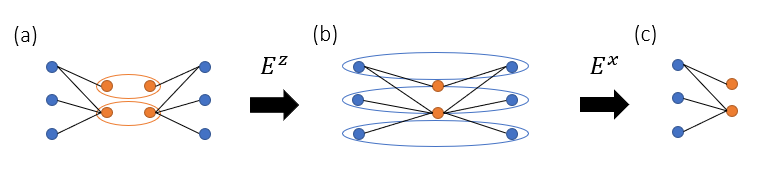}
    \caption{Scheme to extend the Hadamard map to general two-colorable graph states. (a) Taking two copies of the graph one applies $E^z$ to the corresponding pairs of nodes of the same color (orange). (b) Then, the operation $E^x$ is applied to the pairs of nodes of the second color (blue). (c) Finally, we obtain the initial graph state. This shows that for two-colorable graph states there is a Hadamard-type map projecting two-copies to a single copy, where the graph state is a fixed point.}
    \label{fig:hadamard2colorable}
\end{figure}

As we have seen above, the two projections $E^x$ and $E^z$ 
fuse two nodes
while keeping all connections to common or unequal neighbours, respectively. In this 
section we describe how these two operations can therefore be used to map 
two copies (and therefore iteratively any number of copies) of two-colorable 
graphs to one copy of the same graph.
This implies that the method of the Hadamard map can be extended
from GHZ states to all two-colorable graph states.  

A graph is called two-colorable if one can assign one of two different 
colors to every node, such that nodes with a common edge never have 
the same color. One example is shown in Fig.~\ref{fig:hadamard2colorable}. 
We use the colors orange and blue for simplicity.  

If we now take two copies of the same two-colorable graph, we can map 
it to one copy by the following simple scheme (see also Fig.~\ref{fig:hadamard2colorable}):
\begin{enumerate}[I]
    \item Choose one of the two colors (here: orange) and apply the 
    operation $E^z$ to all orange nodes together with their 
    corresponding nodes in the second copy of the graph.
    
    \item Apply $E^x$ to every blue node and its corresponding 
    node stemming from the second copy of the graph.
\end{enumerate}
The first step uses the fact that two nodes with different neighbours keep these neighbours under the action $E^z$. So after the first step, all blue nodes keep their connections to the orange nodes. Further, all pairs of blue nodes from copy one and copy two have only common neighbours. So, in the second step our graphical rule for $E^x$ can be applied. The projection $E^x$ then keeps these connections to the common neighbours while mapping two blue nodes to one again. The remaining state is then exactly one copy of the original state.

%%%%%%%%%%%%%%%%%%%%%%%%%%%%%%%%%%%%%%%%%%%%%
\section{Detailed discussion of the results for GHZ-diagonal states}
%%%%%%%%%%%%%%%%%%%%%%%%%%%%%%%%%%%%%%%%%%%%%
\label{sec-app-technicalresults}

In this section we consider GHZ-diagonal states $\rho$ with entries $(1,x,y,z,z,y,x,1)$ on the diagonal and antidiagonal where $1\geq x \geq y \geq z \geq 0$. These states are biseparable if and and only if $1\leq x+y+z$~\cite{GuehneSeevinck2010}. Also, they are partition-separable if $1=x$ and $y=z$. To shorten the notation, we 
write $\rho = \chi(1,x,y,z)$ for the normalized states.  

We now implement the methods described in the main text to explore the properties of the $2$-copy state $\chi(1,x,y,z)$ for different $x$, $y$ and $z$, and how these properties change with additional white noise on the single-copy level.  

\subsection{Two copies of GHZ-diagonal states for different levels of noise}

\begin{figure*}[h]
    \centering
    \includegraphics[width=0.2\linewidth]{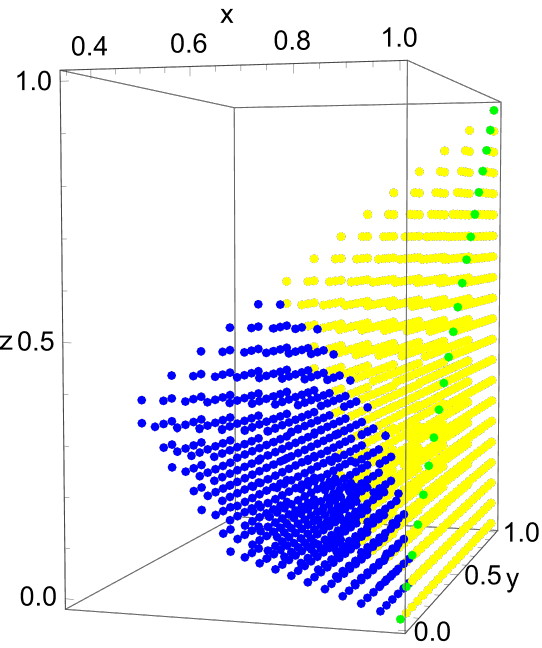}
    \includegraphics[width=0.2\linewidth]{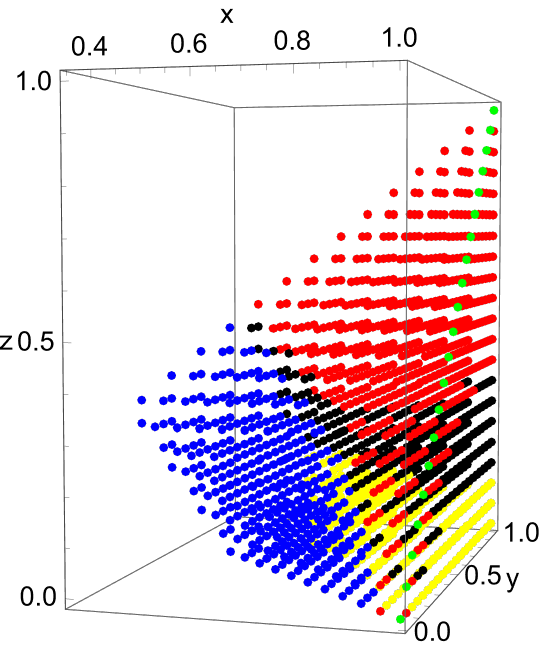}
    \includegraphics[width=0.2\linewidth]{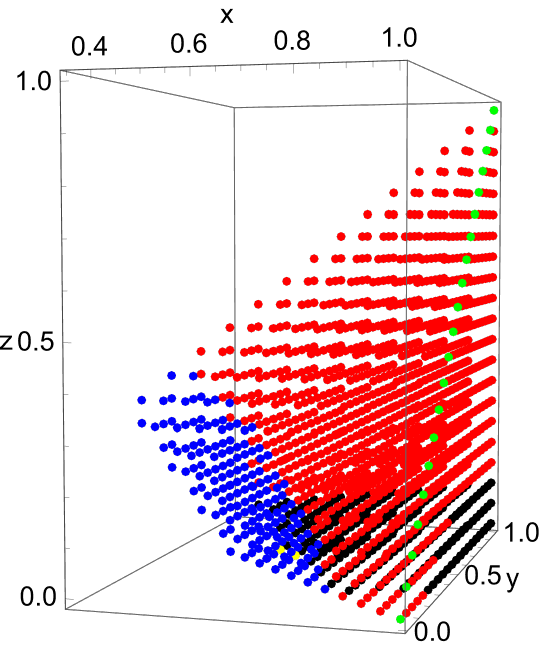}
    \includegraphics[width=0.2\linewidth]{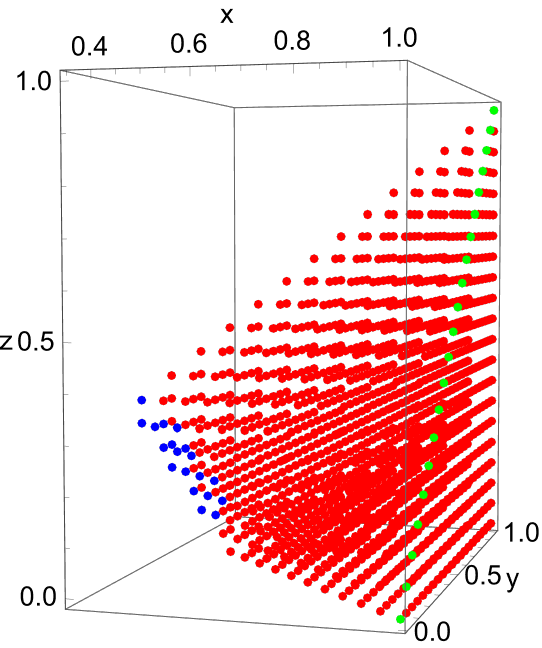}
    \caption{
    Results for GHZ-diagonal states $p\chi(1,x,y,z)+\frac{(1-p)}{8}\one$ for different levels of white noise ($p=0, 0.1, 0.2, 0.3$); the green points are partition separable and therefore not activatable; all but the red points are detected by the PPT mixture approach to be GME; the blue points are detected using the Hadamard map; for the black points it was shown by the PPT relaxation that no projections exist.}\label{fig:3dGHZdiagonal}
\end{figure*}

 We investigate the states
\begin{align}
    \varrho = p \chi(1,x,y,z) + \frac{(1-p)}{8}\one_8
\end{align}
for four different levels of noise, that is $p=1$, $0.9$, $0.8$ and $p=0.7$, and consider numerically every possible assignment of $(x,y,z)$, such that the state $\chi(1,x,y,z)$ is biseparable, in steps of $0.05$. This leads to $1457$ data points. The results are displayed in Fig.~\ref{fig:3dGHZdiagonal}.  

First, we note that the state $\chi(1,x,y,z)$ is partition-separable if $1=x$ and $y=z$ and therefore can never be activated. These points are represented in green.  

For all other states we apply the PPT mixture approach to the $2$-copy state $\varrho^{\otimes 2}$ to check whether they are GME. Interestingly, we find that for no noise (that is, $p=1$) the states $\varrho^{\otimes 2}=\chi(1,x,y,z)^{\otimes 2}$ are always detected as GME. If the level of noise increases there are however states which are not detected anymore as GME, displayed in red.  

Applying the Hadamard map to the states $\varrho^{\otimes 2}$ which are GME we find that in the case of up to $p=0.8$ around $40\%$ of the considered GME states are detected by the Hadamard map. In the case of $p=0.7$, however, the method of PPT mixtures and the Hadamard map detect exactly the same states. The states detected by the Hadamard map are displayed in blue.  

The remaining states have now the properties that they are detected using the PPT mixture approach to be GME but their entanglement can not be detected by the Hadamard map. Using the PPT relaxation
described in Appendix~\ref{sec-app-finding-projections}
we find that some of these states can never be detected using the scheme of local projections and the GHZ-fidelity witness
\begin{align}
    \WW_{\text{GHZ}} = \frac{1}{2}\one_8 - \ketbra{\mathrm{GHZ_+}}{\mathrm{GHZ_+}}.
\end{align}
These points are displayed in black. For the remaining yellow points there could exist projections which detect the entanglement, but the seesaw algorithm does not always find some. In the case of no noise (that is, $p=1$) the seesaw algorithm finds projections for every yellow point, so, every state $\chi(1,x,y,z)$ which is not partition-separable can actually be detected by local projections and the GHZ-fidelity witness. However, in the case of $p=0.9$ there are some states for which the seesaw algorithm finds no projection leading to detection. It is unclear whether this is a numerical property or can also be shown analytically, see also the discussion in  
Appendix \ref{sec-app-finding-projections}. 

\subsection{White noise robustness of the $2$-copy GHZ-diagonal state}

For the white-noise robustness of $\rho^{\otimes 2}$, that is the value $p_{\text{wnr}}$ for which $p\rho^{\otimes 2} + \frac{1-p}{64}\one_{64}$ becomes a PPT mixture, one has 
\begin{align}
    \tr[(p_{\text{wnr}} \rho^{\otimes 2} + \frac{1-p_{\text{wnr}}}{64} \one_{64}) \WW] = 0
\end{align}
for the witness $\WW$ found by the PPT mixture approach. Using the result $\tr[\rho \WW] = t$ coming from the SDP and the fact that the witness $\WW$ is normalized $\tr[\WW]=1$, one therefore finds 
\begin{align}
    p_{\text{wnr}} = \frac{1}{1-64 t}.
\end{align}
Note that $\rho^{\otimes 2}$ is GME for all $p\geq p_{\text{wnr}}$. In Fig.~3 in the main text this value is displayed for the GHZ-diagonal states considered above. One finds that the most robust GHZ-diagonal state is the state $\chi(1,1/3,1/3,1/3)$ with a noise robustness of $p_{\text{wnr}} = 0.5294$. 

\subsection{GHZ fidelity after local projections of two copies}

We are interested in the GHZ fidelity that the state $\varrho^{\otimes 2}$ can reach after applying local projections to it, that is in the quantity $F_{\mathrm{GHZ}}(\tau_2) = \tr(\tau_2  \ketbra{\mathrm{GHZ}}{\mathrm{GHZ}})$ where $\tau_2$ is the projected and renormalized state.   

As explained in the main text, the overlap of the GHZ state and the state $\rho^{\otimes 2}$ directly after the application of the local projections $\ket{a}$, $\ket{b}$ and $\ket{c}$ can be calculated by
\begin{align}
    f:= \tr(\tilde{\varrho}\ketbra{\mathrm{GHZ}}{\mathrm{GHZ}})
    =\expten{a b c}{\varrho^{\otimes 2}}{\ketbra{\mathrm{GHZ}}{\mathrm{GHZ}}}.
\end{align}
However, here the state $\tilde{\varrho}$ is not yet normalized. As a normalization constant we obtain
\begin{align}
    \mathcal{N} = \tr(\tilde{\varrho}\one_8)
    =\expten{a b c}{\varrho^{\otimes 2}}{\one_8}, 
\end{align}
then the normalized state is $\tau_2 = \frac{1}{\mathcal{N}} \tilde{\varrho}$ and the GHZ fidelity is $F_{\mathrm{GHZ}}(\tau_2) = \frac{f}{\mathcal{N}}$.   

From the numerical calculations above we already obtained projections which minimize the value
\begin{align}
    X &:=\, \expten{a b c}{\varrho^{\otimes 2}}{\WW_{\text{GHZ}}} \,
    =\, \frac{1}{2}\tr{\bigl(\tilde{\varrho}\,\one_8}\bigr) -\tr{\bigl(\tilde{\varrho}\,\ketbra{\mathrm{GHZ}}{\mathrm{GHZ}}\bigr)} 
    \,=\, \frac{\mathcal{N}}{2} - f , 
\end{align}
so we find 
\begin{align}
    F_{\mathrm{GHZ}}(\tau_2) &\geq\, \frac{1}{2} - \frac{X}{\mathcal{N}}.
\end{align}

Note that this method only leads to a lower bound on the accessible fidelity, since we optimized the product vectors with respect to the difference of $\mathcal{N}$ and $f$ and not their fraction. To get a better bound on the fidelity, we use again the seesaw algorithm for the witness $\mathcal{W}_{\mathrm{GHZ}, \kappa} = \kappa \one - \ketbra{\mathrm{GHZ}}{\mathrm{GHZ}}$ and search for the highest $\kappa$ for which $\tr(\tau_2  \mathcal{W}_{\mathrm{GHZ}, \kappa})$ is negative. This leads directly to a lower bound $F_{\mathrm{GHZ}}(\tau_2) \geq \kappa$ on the fidelity.

In Fig.~3 in the main text this value is displayed for the GHZ-diagonal states considered above. One finds that the state with the highest noise robustness $\chi(1,1/3,1/3,1/3)$ has a GHZ fidelity of $0.75$ and lies in the W class. This can be seen from the facts, that this state and the W state $\ket{W} = \frac{1}{\sqrt{3}}(\ket{++-} + \ket{+-+} + \ket{-++})$ have the same GHZ fidelities for the different GHZ states, and the state $\chi(1,1/3,1/3,1/3)$ is a depolarized version of the W state.   

Interestingly, the states with the highest GHZ fidelity can be found for $x\approx 1$, $y\approx 0$ and $z=0$. We found the highest fidelity of $0.97561$ for the state $\chi(1,1,0.05,0)$, which is the state closest to the partition-separable state $\chi(1,1,0,0)$ that we considered in the numerical calculations. One possible set of local projections achieving this value is given by
\begin{subequations}
\begin{align}
    F_A &=\, \ketbra{0}{0+} + \ketbra{1}{1-}, \\[1mm]
    F_B &=\, \ketbra{0}{1+} + \ketbra{1}{0-} \text{ and} \\[1mm]
    F_C &=\, \ketbra{0}{0+} + \ketbra{1}{1+}. 
\end{align}
\end{subequations}
Applying these projections to the state $\chi(1,1,\epsilon,0)$ and using the same notation as in Eq.~\eqref{Xstates} we find
\begin{align}
     \tau_2 &=\, \frac{1}{\mathcal{N}} \left( 2\epsilon\,\ketbra{\mathrm{GHZ}_{+1}}{\mathrm{GHZ}_{+1}} + \epsilon^2\,\ketbra{\mathrm{GHZ}_{-1}}{\mathrm{GHZ}_{-1}} \right) \nonumber\\[1mm]
     &=\, \frac{2}{2+ \epsilon}\,\ketbra{\mathrm{GHZ}_{+1}}{\mathrm{GHZ}_{+1}} + \frac{\epsilon}{2+ \epsilon}\,\ketbra{\mathrm{GHZ}_{-1}}{\mathrm{GHZ}_{-1}} ,
\end{align}
and, more generally, one finds for the state $\chi(1,\mu,\epsilon,\delta)$ the projected state
\begin{align}
    \tau_2 &=\, \frac{1}{\mathcal{N}} \bigl( \epsilon(1+\mu)\,\ketbra{\mathrm{GHZ}_{+1}}{\mathrm{GHZ}_{+1}} + \epsilon(\epsilon+\delta)\,\ketbra{\mathrm{GHZ}_{-1}}{\mathrm{GHZ}_{-1}}\nonumber\\[1mm] 
    &\ \ \ \ \ +\, \delta(1+\mu)\,\ketbra{\mathrm{GHZ}_{+2}}{\mathrm{GHZ}_{+2}} + \delta(\epsilon + \delta)\,\ketbra{\mathrm{GHZ}_{-2}}{\mathrm{GHZ}_{-2}} \bigr),
\end{align}
where $\mathcal{N}$ is the normalization constant. As the entries of $\chi(1,\mu,\epsilon,\delta)$ are in decreasing order this also leads to a criterion for $2$-copy GME activation on the single-copy level: As the local projections can never create entanglement and the projected state is entangled if the largest GHZ fidelity is greater than $0.5$, we find that the state $\chi(1,\mu,\epsilon,\delta)$ is $2$-copy GME if
\begin{align}\label{app-eq:criteriondistill}
    \frac{\epsilon (1+\mu)}{(\epsilon+\delta) (1 + \mu + \epsilon + \delta)} > 0.5.
\end{align}

Note that this criterion detects different states than the Hadamard map. Considering the states with $p=0$ in Fig.~\ref{fig:3dGHZdiagonal} we find that the Hadamard map and the criterion stated above detect each about $40\%$ of the states detected by the PPT mixture approach. However, only $20\%$ of these states are detected by both criteria, while the other $20\%$ are unique to each criterion. This difference can be explained in a more general way. Since the Hadamard map projects two copies of GHZ-diagonal states again on a GHZ-diagonal state, GME can be easily detected in that state by applying, e.g., the criterion in Ref.~\cite{GuehneSeevinck2010}. However, for the criterion above we specifically are interested in the GHZ fidelity after projecting the state, discarding the property of preserving the diagonal form during the projection. This can be better understood in the context of entanglement distillation. Similar to distillation protocols, the biseparable state has in the beginning a low GHZ fidelity. Taking two copies and performing local projections corresponds to a measurement of the local $2$-qubit systems and post-selecting on the desired output state.

%%%%%%%%%%%%%%%%%%%%%%%%%%%%%%%%%%%%%%%%%%%%%%%%%%%%%%%%%%%%%%%%%%%%%
\section{Multi-copy GME criterion evaluated on single-copy density-matrix elements}
\label{app-sec:GabrielWitness}
%%%%%%%%%%%%%%%%%%%%%%%%%%%%%%%%%%%%%%%%%%%%%%%%%%%%%%%%%%%%%%%%%%%%%

 In this Appendix, we present and discuss a generalization of a criterion 
for the detection of GME derived by \emph{Gabriel}, \emph{Hiesmayr}, and 
\emph{Huber} (GHH) in Ref.~\cite{GabrielHiesmayrHuber2010} to the multi-copy 
case, which we dub the \emph{$k$-copy GHH criterion}. In 
Sec.~\ref{appendix:the k-copy GHH criterion} we present a derivation 
of the $k$-copy GHH criterion, before discussing the feasibility of 
testing this criterion experimentally relative to general GME tests 
on the multi-copy Hilbert space and the relation to the single-copy 
criterion in more detail. In Sec.~\ref{subsec:k-copy GHH criterion for GHZ-diagonal states} 
we then apply the $k$-copy GHH criterion to the special case of GHZ-diagonal states. 
%%%%%%%%%%%%%%%%%%%%%%%%%%%%%%%%%%%%%%%%%%%%%%%%%%%%%%%%%%%%%%%%%%%%%

\subsection{The $k$-copy GHH criterion}\label{appendix:the k-copy GHH criterion}

%%%%%%%%%%%%%%%%%%%%%%%%%%%%%%%%%%%%%%%%%%%%%%%%%%%%%%%%%%%%%%%%%%%%%

We now present a $k$-copy GME witness, which is constructed as a $k$-copy 
version of an $m$-separability criterion from Ref.~\cite{GabrielHiesmayrHuber2010}. 
Referring to the original criterion in~\cite{GabrielHiesmayrHuber2010} as the 
\emph{GHH criterion}, we call the generalization we derive here the 
\textit{$k$-copy GHH criterion}. The original GHH criterion states that 
every $m$-separable $N$-partite state $\rho$ satisfies
\begin{align}\label{app-eq:GabrielWitness}
    \sqrt{ \brakket{\Phi}{\rho^{\otimes 2}P_{\text{tot}}}{\Phi} } 
    \le 
    \sum\limits_{\{\alpha\}} \bigg( \prod\limits_{i=1}^m \brakket{\Phi}{P_{\alpha_i}^\dagger\rho^{\otimes 2}P_{\alpha_i}}{\Phi} \bigg)^\frac{1}{2m}
\end{align}
for all pure fully separable $2N$-partite states 
$\ket{\Phi}=\bigotimes_{i=1}^{2N}\ket{\Phi\suptiny{0}{0}{(i)}}$, where 
the sum runs over all possible partitions $\alpha$ of the considered system 
into $m$ subsystems, the permutation operators $P_{\alpha_i}$ are the operators 
permuting the two copies of all subsystems contained in the $i$-th subset 
of the partition $\alpha$ and $P_{\text{tot}}$ is the total permutation operator, 
permuting the two copies. As the inequality in~(\ref{app-eq:GabrielWitness}) is 
satisfied for all $m$-separable states, a violation for $m=2$ witnesses GME in the 
state~$\rho$.  

To derive a version of the GHH criterion in Eq.~(\ref{app-eq:GabrielWitness}) for 
a $k$-copy state $\tilde{\rho}=\rho^{\otimes k}$, we start by noting that the 
fully separable state $\ket{\Phi}$ will now be of the form 
$\ket{\tilde{\Phi}}= \ket{A}\otimes \ket{B} = \ket{a_1\dots a_k} \otimes \ket{b_1\dots b_k}$, 
where $\ket{A}$ and $\ket{B}$ are each states on the Hilbert spaces 
corresponding to two copies $\tilde{\rho}^{\otimes 2}$, and 
$\ket{a_n}=\bigotimes_{i=1}^N \ket{a_{n_i}}$ and 
$\ket{b_n}=\bigotimes_{i=1}^N \ket{b_{n_i}}$ are $N$-partite 
fully separable states. Hence, if we denote the fully separable state for the single-copy GHH criterion  as $\ket{\tilde{\Phi}_n} = \ket{a_n b_n}$, 
then $\ket{\tilde{\Phi}}$ satisfies 
$\ket{\tilde{\Phi}} = \ket{\tilde{\Phi}_1} \otimes \dots \otimes \ket{\tilde{\Phi}_k}$. 
Further, since 
$P \ket{\tilde{\Phi}} = P (\ket{a_1\dots a_k} \otimes \ket{b_1\dots b_k}) = (P \ket{\tilde{\Phi}_1}) \otimes \dots \otimes (P \ket{\tilde{\Phi}_k})$, the 
left-hand side of Eq.~(\ref{app-eq:GabrielWitness}) takes the form
\begin{align}
    \sqrt{\brakket{\tilde{\Phi}}{\tilde{\rho}^{\otimes 2} P_{\text{tot}}}{\tilde{\Phi}}}
    = \prod_{n=1}^k \sqrt{\brakket{\tilde{\Phi}_n}{\rho^{\otimes 2} P_{\text{tot}}}{\tilde{\Phi}_n}}\,,
\end{align}
while the right-hand side can be expressed as
\begin{align}
    \sum\limits_{\{\alpha\}} \left( \prod\limits_{i=1}^2 \brakket{\tilde{\Phi}}{P_{\alpha_i}^\dagger\tilde{\rho}^{\otimes 2}P_{\alpha_i}}{\tilde{\Phi}} \right)^\frac{1}{4}
    = \sum\limits_{\{\alpha\}} \left( \prod\limits_{i=1}^2 \prod\limits_{n=1}^k \brakket{\tilde{\Phi}_n}{P_{\alpha_i}^\dagger\rho^{\otimes 2}P_{\alpha_i}}{\tilde{\Phi}_n} \right)^\frac{1}{4}.
\end{align}

Thus, we can formulate the $k$-copy version of~(\ref{app-eq:GabrielWitness}) as follows: Any state $\rho$ for which the $k$-copy state $\tilde{\rho}=\rho^{\otimes k}$ is biseparable satisfies 
\begin{align}\label{app-eq:kcopyGabrielWitness}
    \prod_{n=1}^k \sqrt{\brakket{\Phi_n}{\rho^{\otimes 2} P_{\text{tot}}}{\Phi_n}} 
    \leq 
    \sum\limits_{\{\alpha\}} \left( \prod\limits_{i=1}^2 \prod\limits_{n=1}^k \brakket{\Phi_n}{P_{\alpha_i}^\dagger\rho^{\otimes 2}P_{\alpha_i}}{\Phi_n} \right)^\frac{1}{4}
\end{align}
for all fully separable states $\ket{\Phi_n}$, with $n=1,\dots,k$.  

One may note here that, since a partition $\alpha$ is defined as a grouping of the set of parties $[N]:=\{1,\cdots, N\}$ into disjoint subsets whose union is equal to $N$, it can be seen from Eq.~(\ref{app-eq:kcopyGabrielWitness}) that for an $N$-partite state $\rho$, the biseparability of the $k$-copy state $\rho^{\otimes k}$ refers to the biseparability of an $N$-partite state, where the $j$th party is understood as a set of $j$th parties of all the copies of $\rho^{\otimes k}$, i.e., $j=\{j_1,\dots j_k\}$. For instance, for three parties ($N=3$) labelled $A$, $B$, and $C$, the subsystems of two copies ($k=2$) would be $A_1$, $B_1$, and $C_1$ for the first copy, along with $A_2$, $B_2$, and $C_2$ for the second copy, and bipartition $A|BC$ of the two-copy state would hence refer to the bipartition $A_1A_2|B_1B_2C_1C_2$.  

%%%%%%%%%%%%%%%%%%%%%%%%%%%%%%%%%%%%%%%%%%%%%%%%%%%%%%%%%%%%%%%%%%%%%%%%%%%

As already explained in Ref.~\cite{GabrielHiesmayrHuber2010}, using the original 
GHH witness in~(\ref{app-eq:GabrielWitness}) does not require full state tomography 
of $\rho$, since only knowledge of certain matrix elements of $\rho$ is required for 
a fixed $\ket{\Phi}$, which could in principle be obtained from fewer measurements 
than would be necessary for state tomography. Further, as we discussed in the Section~"Experimental detection of superactivation" in the main text, although at first sight the detection of superactivation requires joint measurements across subsystems pertaining to different copies, entanglement witnesses (linear or nonlinear) on a multi-copy Hilbert space can be decomposed into tensor products of observables 
on single parties. This leads to a nonlinear expression on the single-copy level, which
can be evaluated on a single copy using local measurements. In case of the GHH criterion, this can be seen from Eq.~(\ref{app-eq:kcopyGabrielWitness}) where taking into account that all copies are identical, evaluating the witness for $k$ copies comes down to measuring the same matrix elements as for the single-copy case and taking the $k$-th powers of each of them, meaning that any experimental tests of the $k$-copy GHH criterion would be exactly as (in)expensive as testing the original single-copy criterion.  

Another way to evaluate GME criteria on the $k$-copy state can be achieved by mapping the $k$-copy state to a single-copy state. To do this, one can use the approach presented in Ref.~\cite{YamasakiMorelliMiethlingerBavarescoFriisHuber2022}, which employs the Hadamard map $\mathcal{E}$ that maps the tensor product of two density operators $\rho$ and $\sigma$ both defined on the Hilbert space $\mathcal{H}$ (such that their tensor product $\rho\otimes\sigma$ is defined on $\mathcal{H}^{\otimes 2}$), {so the resulting normalized state on $\mathcal{H}$ reads}
\begin{align}\label{app-eq:Hadamardmap}
    \mathcal{E}(\rho\otimes\sigma)
    =
    \frac{\rho\circ\sigma}{\tr{(\rho\circ\sigma)}}\quad \text{on}\quad \mathcal{H},
\end{align}
where $``\circ"$ denotes the Hadamard (or Schur) product, i.e., component-wise multiplication of all matrix elements. This map can, as explained in the main text, be implemented via stochastic local operations and classical communication (SLOCC)~\cite{LamiHuber2016}, and hence cannot create entanglement where none was present before. However, one should note that this map may lead to a reduction of entanglement.  

As an example how this projection might be used, we consider the state $\rho_\mathrm{act}=\bigl(\ketbra{\mathrm{GHZ}_+}{\mathrm{GHZ}_+} +\rho_{\mathrm{sym}}\bigr)/2$, which is a biseparable GHZ-symmetric state and lies on the boundary to the GME states as well as on the boundary of the GHZ-diagonal states. 
The distance of the projected $k$-copy state to the state $\ket{\mathrm{GHZ}_+}$ diminishes with the number of copies $k$ as
\begin{align}
    \tr\bigl(\ketbra{\mathrm{GHZ}_+}{\mathrm{GHZ}_+}\rho_\mathrm{act}^{\circ k}\bigr) &=\,\frac{3^{k-1}}{3^{k-1}+1},
    \label{eq:iteration}
\end{align}
where $\rho^{\circ k}$ corresponds to the $k$-wise Schur product of the state $\varrho$.
We have thus found a sequence that approaches the GHZ state. The projected $k$-copy states are also shown as blue dots in Fig.~4 of the main text.
Note that the projected $2$-copy state already lies on the boundary of the GHZ class, implying GME.

Now, we show that reducing $\rho^{\otimes k}$ to a single copy using the Hadamard map and then applying the original GHH criterion~(\ref{app-eq:GabrielWitness}) is equivalent to using the $k$-copy GHH criterion~(\ref{app-eq:kcopyGabrielWitness}). 
To this end we define the state $\mathcal{E}_{k-1}(\rho^{\otimes k})$ obtained from successively applying the Hadamard map from Eq.~(\ref{app-eq:Hadamardmap}) $k-1$ times to $k$ copies $\rho_1=\rho_2=\ldots=\rho_k=\rho$ of a single-copy state $\rho$, i.e.,
\begin{align}\label{app-eq:rhocirc}
    \mathcal{E}_{k-1}(\rho^{\otimes k})
    =\mathcal{E}_\circ[\rho_1\otimes\mathcal{E}_\circ[\rho_2\otimes\mathcal{E}_\circ[...\otimes\mathcal{E}_\circ[\rho_{k-1}\otimes\rho_k]]]]
    \,=\,\frac{\rho_1\circ\cdots\circ\rho_k}{\tr{(\rho_1\circ\cdots\circ\rho_k)}}
    \,=\,\frac{\rho^{\circ k}}{\tr{(\rho^{\circ k})}}\,.
\end{align}
Then, by substituting Eq.~(\ref{app-eq:rhocirc}) into both sides of the inequality in~(\ref{app-eq:GabrielWitness}) and taking into account that $\brakket{\Phi}{\rho^{\circ k}}{\Phi}=\prod_{n=1}^k\brakket{\Phi_n}{\rho}{\Phi_n}$, we get for the left-hand side
\begin{align}
   \sqrt{ \brakket{\Phi}{
   \mathcal{E}_{k-1}(\rho^{\otimes k})
   ^{\otimes 2}P_{\text{tot}}}{\Phi} }  =
   \sqrt{\frac{1}{[\tr{(\rho^{\circ k})}]^2}\brakket{\Phi}{(\rho_1\circ\cdots\circ\rho_k)^{\otimes 2}P_{\text{tot}}}{\Phi}}=
   \frac{1}{\tr{(\rho^{\circ k})}}\prod_{n=1}^k\sqrt{\brakket{\Phi_n}{\rho^{\otimes 2}P_{\text{tot}}}{\Phi_n}}
\end{align}
and for the right-hand side we get
\begin{subequations}
\begin{align}
    \sum\limits_{\{\alpha\}} \bigg( \prod\limits_{i=1}^2 \brakket{\Phi}{P_{\alpha_i}^\dagger
    \mathcal{E}_{k-1}(\rho^{\otimes k})
    ^{\otimes 2}P_{\alpha_i}}{\Phi} \bigg)^\frac{1}{4}
    & =\, \sum\limits_{\{\alpha\}} \bigg( \prod\limits_{i=1}^2 \frac{1}{[\tr{(\rho^{\circ k})}]^2}\brakket{\Phi}{P_{\alpha_i}^\dagger(\rho_1\circ\cdots\circ\rho_k)^{\otimes 2}P_{\alpha_i}}{\Phi} \bigg)^\frac{1}{4} \\
    & =\,  \frac{1}{\tr{(\rho^{\circ k})}} \sum\limits_{\{\alpha\}} \bigg( \prod\limits_{i=1}^2 \prod\limits_{n=1}^k\brakket{\Phi_n}{P_{\alpha_i}^\dagger\rho^{\otimes 2}P_{\alpha_i}}{\Phi_n} \bigg)^\frac{1}{4}.\nonumber
\end{align}
\end{subequations}
One can clearly see that the normalization factor $1/\tr{(\rho^{\circ k})}$ on the left-hand side and right-hand side cancels out, and choosing 
$\ket{\Phi}=\bigotimes_{n=1}^{k}\ket{\Phi_n}$, we recover the original GHH criterion for the $(k-1)$-fold Hadamard product exactly in the form~(\ref{app-eq:kcopyGabrielWitness}).  

Hence, we showed that using the $k$-copy GHH criterion~(\ref{app-eq:kcopyGabrielWitness}) is equivalent to mapping the $k$-copy state $\rho^{\otimes k}$ to a single copy using the Hadamard map and applying the original single-copy version of the GHH criterion~(\ref{app-eq:GabrielWitness}).

%%%%%%%%%%%%%%%%%%%%%%%%%%%%%%%%%%%%%%%%%%%%%%%%%%%%%%%%%%%%%%%

\subsection{The $k$-copy GHH criterion for GHZ-diagonal states}\label{subsec:k-copy GHH criterion for GHZ-diagonal states}

%%%%%%%%%%%%%%%%%%%%%%%%%%%%%%%%%%%%%%%%%%%%%%%%%%%%%%%%%%%%%%%

The criterion~(\ref{app-eq:kcopyGabrielWitness}) can be simplified even further when restricting to particular families of states. Here, we consider so-called GHZ-diagonal states, which are convex combinations $\rho_{\mathrm{GHZ}} = \sum_{n} p_n \ketbra{\mathrm{GHZ}_n}{\mathrm{GHZ}_n}$ of projectors onto the $8$ GHZ-type states defined as
\begin{subequations}
\label{Xstates}
\begin{align}
    \ket{\mathrm{GHZ}_{\pm 1}} &=\,\tfrac{1}{\sqrt{2}}\bigl( \ket{000} \pm \ket{111}\bigr), \\
    \ket{\mathrm{GHZ}_{\pm 2}} &=\,\tfrac{1}{\sqrt{2}}\bigl( \ket{001} \pm \ket{110}\bigr), \\
    \ket{\mathrm{GHZ}_{\pm 3}} &=\,\tfrac{1}{\sqrt{2}}\bigl( \ket{010} \pm \ket{101}\bigr), \\
    \ket{\mathrm{GHZ}_{\pm 4}} &=\,\tfrac{1}{\sqrt{2}}\bigl( \ket{100} \pm \ket{011}\bigr)
\end{align}
\end{subequations}
with $\sum_n p_n=1$. 
For our witness, we need to consider fully separable $2N$-partite pure states $\ket{\Phi_n}$, and here $N=3$, but for the choice of GHZ-diagonal states we restrict our considerations to four possible choices for each $\ket{\Phi_n}$ to be 
\begin{subequations}\label{eq:vector choices}
\begin{align}
    \ket{\Psi_1} &=\, 
    \ket{000 111}, \\
    \ket{\Psi_2} &=\, 
    \ket{001 110}, \\
    \ket{\Psi_3} &=\, 
    \ket{010 101}, \\
    \ket{\Psi_4} &=\, 
    \ket{100 011}.
\end{align}
\end{subequations}
Now, for the $k$-copy GHH criterion the state $\ket{\Phi}$ is a tensor product of $k$ such vectors, where each of the four vectors $\ket{\Psi_m}$ in Eq.~(\ref{eq:vector choices}) for $m=1,2,3,4$ occurs $k_m$ times and the $k_m$ sum up to $k$. Straightforward calculation of the witness then leads to the following statement:
Every GHZ-diagonal state $\rho_{\mathrm{GHZ}}$, for which $k$ copies are biseparable, satisfies  
\begin{align}\label{eq:witnessGZHdiagonal}
    \prod_{i=1}^4 \Lambda_i^{k_i} \leq 
    M_4^{k_1}M_3^{k_2}M_2^{k_3}M_1^{k_4} + 
    M_3^{k_1}M_4^{k_2}M_1^{k_3}M_2^{k_4} + 
    M_2^{k_1}M_1^{k_2}M_4^{k_3}M_3^{k_4}
\end{align}
for all combinations of $0\leq k_m \leq k$ with $\sum_{m=1}^4 k_m = k$, where $\Lambda_m = |\lambda_{+m} - \lambda_{-m} |$ and $M_m = (\lambda_{+m} + \lambda_{-m})$, and $\lambda_{\pm m}$ are the eigenvalues of $\rho_{\mathrm{GHZ}}$ corresponding to the eigenstates $\ket{\mathrm{GHZ}_{\pm m}}$.  

%%%%%%%%%%%%%%%%%%%%%%%%%%%%%%%%%%%%%%%%%%%%%%%%%%%%%%

\begin{figure}[h]
    \centering
    \large{(a)}\includegraphics[width=0.42\textwidth]{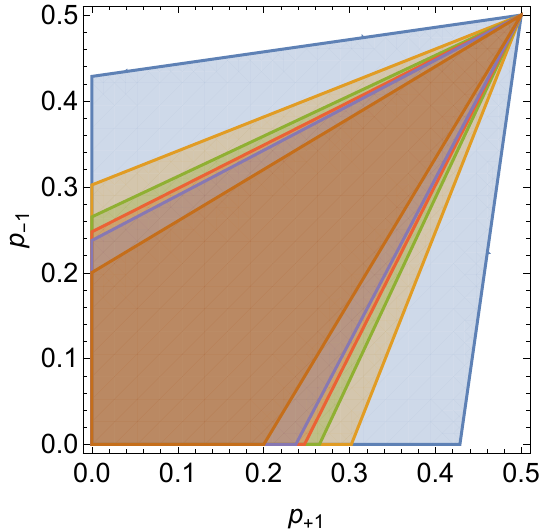}
    \ \ \ \ \large{(b)}\includegraphics[width=0.42\textwidth]{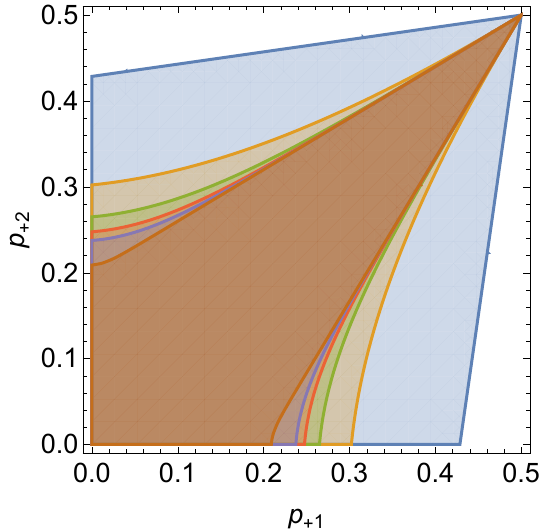}
    \caption{\textbf{(a)} Regions of $p_{+1}$ and $p_{-1}$ where the state $\rho_1$ is \textit{not} detected as $k$-copy activatable for $k=1,2,3,4,5$ and $k=1000$; the region gets smaller with increasing $k$, i.e., the largest blue region corresponds to the states that are not detected as GME for 2 copies, and it contains all the remaining regions corresponding to larger $k$. 
    \textbf{(b)} Regions of $p_{+1}$ and $p_{+2}$ where the state $\rho_2$ is \textit{not} detected as $k$-copy activatable for $k=2,3,4,5$ and $k=20$; the region gets smaller with increasing $k$, i.e., the largest blue region corresponds to the states that are not detected as GME for $1$ copy, and it contains all the remaining regions corresponding to larger $k$.}
    \label{fig:GHZ+1-1}
\end{figure}

Let us now apply the above result to the specific GHZ-diagonal states
defined as GHZ-symmetric states in the main text. In fact, we are considering
a subset of the state space depicted in Fig.~4 in
the main text, namely the states $\rho_{\mathrm{GHZ}_1}$ defined as a 
convex combination of a GHZ-diagonal state with only two non-zero 
probability weights $p_{+1}$ and $p_{-1}$ and white noise, i.e., 
\begin{align}\label{eq:rho1}
    \rho_1 = p_{+1}\ketbra{\mathrm{GHZ}_{+1}}{\mathrm{GHZ}_{+1}} + p_{-1}\ketbra{\mathrm{GHZ}_{-1}}{\mathrm{GHZ}_{-1}} + \tfrac{1}{8}(1-p_{+1}-p_{-1})\,\mathds{1}.
\end{align}
The eigenvalues $\lambda_i$ of $\rho_1$ are given by $\lambda_{+1} = p_{+1} + \frac{1-p_{+1}-p_{-1}}{8}$, $\lambda_{-1} = p_{-1} + \frac{1-p_{+1}-p_{-1}}{8}$ and $\lambda_{\pm i} = \frac{1-p_{+1}-p_{-1}}{8}$ for all $i\neq 1$. Using Eq.~\eqref{eq:witnessGZHdiagonal} we directly get that $\rho_1^{\otimes k}$ is detected to be GME, if 
\begin{align}
    |\,p_{+1} -p_{-1}|^k > 3 \left( \frac{1-p_{+1}-p_{-1}}{4} \right)^k.
\end{align}
In the case that $p_{+1} > p_{-1}$, 
the previous condition can be rewritten as
\begin{align}\label{eq:rho1crit}
    p_{+1} > \frac{\sqrt[k]{3} }{4+\sqrt[k]{3}} + \frac{4-\sqrt[k]{3} }{4+\sqrt[k]{3}} \,p_{-1}\,.
\end{align}
For $k\rightarrow \infty$ this condition converges to $p_{+1} > \frac{1}{5} + \frac{3}{5} p_{-1}$. This coincides with a threshold for detecting bipartite entanglement for individual bipartitions by the PPT criterion~\cite{HorodeckiMPR1996}, i.e., here, detection of activatability of the states according to the $k$-copy GHH criterion coincides with the boundary
of partition separability.
The regions of $p_{+1}$ and $p_{-1}$ that were not detected by the criterion in Eq.~(\ref{eq:rho1crit}) as GME for $k=1,2,3,4,5$ and $k=1000$ are depicted in Fig.~\ref{fig:GHZ+1-1}~(a).

Having applied the $k$-copy GHH criterion from~\eqref{eq:witnessGZHdiagonal} to states that contain a combination of two GHZ-type states $\ket{\mathrm{GHZ}_{\pm 1}}$
differing only in the relative sign of their index, we now apply the criterion to two GHZ-type states $\ket{\mathrm{GHZ}_{\pm m}}$ that do not differ  
in the sign but have different indices $m$.

In other words, now we can choose a particular noisy GHZ-diagonal state so that it is a convex combination of a GHZ-diagonal state with only two non-zero probability weights $p_{+1}$ and $p_{+2}$ and white noise, i.e.,
\begin{align}\label{eq:rho2}
    \rho_2 &=\, p_{+1}\ketbra{\mathrm{GHZ}_{+1}}{\mathrm{GHZ}_{+1}} + p_{+2}\ketbra{\mathrm{GHZ}_{+2}}{\mathrm{GHZ}_{+2}} + \tfrac{1}{8}(1-p_{+1}-p_{+2})\,\mathds{1}
\end{align} 
Now, the eigenvalues $\lambda_i$ are given by $\lambda_{+1} = p_{+1} + \frac{1-p_{+1}-p_{+2}}{8}$, $\lambda_{+2} = p_{+2} + \frac{1-p_{+1}-p_{+2}}{8}$ and $\lambda_{-1} = \lambda_{-2} = \lambda_{\pm i} = \frac{1-p_{+1}-p_{+2}}{8}$ for all $i\neq 1,2$. Using Eq.~\eqref{eq:witnessGZHdiagonal} we directly get that $\rho_2^{\otimes k}$ is detected to be GME, if there exists $k_1$, such that
\begin{align}
    p_{+1}^{k_1}p_{+2}^{k_2} - (\mu+p_{+2})^{k_1} (\mu +p_{+1})^{k_2} 
    >
    2\mu^k ,
\end{align} 
where $\mu = \frac{1-p_{+1}-p_{+2}}{4}$.  

In the case of $p_{+1} > p_{+2}$ the left-hand side is maximized for $k_1 = k$ 
and $k_2=0$,
such that we have
\begin{align}\label{eq:rho2crit}
    p_{+1}^{k} - (\mu+p_{+2})^{k}
    >
    2\mu^k .
\end{align} 
Once again, the regions of $p_{+1}$ and $p_{+2}$ that were not detected by criterion~(\ref{eq:rho2crit}) as GME for $k=1,2,3,4,5$ and $k=20$ are depicted in Fig.~\ref{fig:GHZ+1-1}~(b). 
In this case, one can clearly see that the dependence of $p_{+1}$ and $p_{+2}$ is nonlinear, in contrast to the case of $\rho_1$. However, in the limit of $k\rightarrow\infty$, it reduces to $5p_{+1}>1+3p_{+2}$, which is linear, and again it coincides with the boundary of partition separability obtained by the PPT criterion.

An interesting observation is that we can obtain the same results for $\rho_1$ and $\rho_2$ by using exactly the same method to detect GME activation as in~\cite{YamasakiMorelliMiethlingerBavarescoFriisHuber2022}, which is based on combining the Hadamard map~(\ref{app-eq:Hadamardmap}) and the GME concurrence~\cite{HashemiRafsanjaniHuberBroadbentEberly2012,MaChenChenSpenglerGabrielHuber2011}.

%%%%%%%%%%%%%%%%%%%%%%%%%%%%%%%%%%%%%%%%%%%%%%%%%%%%%%%%%%%%%%%%%%%%%
\section{Nonlinear witnesses certifying $k$-copy activatability} 
\label{app-sec:FidelityWitness}
%%%%%%%%%%%%%%%%%%%%%%%%%%%%%%%%%%%%%%%%%%%%%%%%%%%%%%%%%%%%%%%%%%%%%

In this section, we describe the procedure of lifting entanglement witnesses
acting on the single-copy level to the multi-copy regime in detail.
This lifting can, for instance, be done by an inversion of the Hadamard 
map. The witness on the multi-copy space is then decomposed into operators
acting locally on each qubit. This can then be interpreted and measured
as a nonlinear witness on a single copy of the state. 
Additionally, we 
demonstrate how one could in principle use this approach in an experiment.

\subsection{Theoretical considerations}

Let us start by considering fidelity witnesses~\cite{BourennaneEiblKurtsieferGaertnerWeinfurterGühneHyllusBrußLewensteinSanpera2004}. Since it is well-known that for GHZ states the fidelity bound is optimal~\cite{GuehneSeevinck2010}, the GHZ-fidelity witness
\begin{align}
    \mathcal{W}_{\mathrm{GHZ}} = \frac{\one}{2} -\ketbra{\mathrm{GHZ_+}}{\mathrm{GHZ_+}} \label{eq:fidWit}
\end{align}
is a good candidate to begin with, where we take $\ket{\mathrm{GHZ_+}} = \ket{\mathrm{GHZ}_{+1}}$ as the standard GHZ state.

In the following, we consider two copies of three-qubit states. However, later we will see that the generalization to $k$ copies of states of $N$ parties is straightforward.
In order to lift this witness to the two-copy space, we apply the dual of the (unnormalized) Hadamard map $\mathcal{E}$.  
We recall that the action of the Hadamard map on one party $X = X_1X_2$ in the two-copy space can be written in the $Z$ basis as
\begin{align}
E_X^z &=\, \ketbra{0}{00} + \ketbra{1}{11}\,,
\end{align}
and accordingly a possible inverse action describing the dual is:
\begin{align}
(E_X^z)^{-1} &=\, \ketbra{00}{0} + \ketbra{11}{1}\,,
\end{align}
where $X =A,B,C$. The index $z$ denotes the basis, however, for simplicity we will omit this index in the following.
Now, we apply the inverse Hadamard map to the single-copy GHZ-fidelity witness Eq.~(\ref{eq:fidWit}):
\begin{align}
    \mathcal{W}_2 &=\, (E_A^{-1} \otimes E_B^{-1} \otimes E_C^{-1})\mathcal{W}_{\mathrm{GHZ}} (E_A^{-1} \otimes E_B^{-1} \otimes E_C^{-1})^\dagger\nonumber\\[1mm]
    &=\, \frac{1}{2}\sum_{i,j,k=0}^1(E_{\{A,B,C\}}^{-1})^{\otimes 3} \ketbra{ijk}{ijk}^{ABC}((E_{\{A,B,C\}}^{-1})^\dagger)^{\otimes 3} - (E^{-1}_{\{A,B,C\}})^{\otimes 3} \ketbra{\mathrm{GHZ_+}}{\mathrm{GHZ_+}}^{ABC} ((E^{-1}_{\{A,B,C\}})^\dagger)^{\otimes 3}\\[1mm]
    \begin{split}
    &=\, \frac{1}{2}  \sum_{i,j,k=0}^1 \ketbra{iijjkk}{iijjkk}^{A_1A_2B_1B_2C_1C_2}-
    \frac{1}{2}( (\ketbra{0}{0}^{\otimes 6})^{A_1A_2B_1B_2C_1C_2} + (\ketbra{1}{1}^{\otimes 6})^{A_1A_2B_1B_2C_1C_2} +\quad\nonumber\\[1mm]
    &\ \ \ + (\ketbra{0}{1}^{\otimes 6})^{A_1A_2B_1B_2C_1C_2} + (\ketbra{1}{0}^{\otimes 6})^{A_1A_2B_1B_2C_1C_2} )
    \end{split} \nonumber\\[1mm]
    &=\, \frac{1}{2}\Bigl(\sum_{\substack{i,j,k=0\\ 0<i+j+k<3}}^1 (\ketbra{ijk}{ijk}^{ABC})^{\otimes 2} - \chi_6\Bigr),\nonumber 
\end{align}
where we have defined $\chi_6 \coloneq \ketbra{0}{1}^{\otimes 6}+\ketbra{1}{0}^{\otimes 6}$ in the last step and the condition $0<i+j+k<1$ in the sum excludes the cases $\ket{ijk}=\ket{000}$ and $\ket{ijk}=\ket{111}$, as these cancel out with diagonal terms of the GHZ state.

Further, we note that the term $\chi_6$ can be decomposed into six terms $\mathcal{M}_l^{\otimes 6}$ which are linear combinations of Pauli-$X$ and $Y$ measurements, or more generally for $n$ qubits~\cite{GuehneLuGaoPan2007}
\begin{align}
    \chi_n = \ketbra{0}{1}^{\otimes n}+\ketbra{1}{0}^{\otimes n} = \sum_{l=1}^n\frac{(-1)^l}{n}\mathcal{M}_l^{\otimes n},
    \label{eq:M}
\end{align}
with $\mathcal{M}_l = \cos(l\pi/n)X+\sin(l\pi/n)Y$ for $l = 1,\ldots,n$. 
Using this for $n=6$, we obtain
\begin{align}
    \mathcal{W}_2 &=\, \frac{1}{2}\Bigl(\sum_{\substack{i,j,k=0\\0<i+j+k<3}}^1(\ketbra{ijk}{ijk}^{ABC})^{\otimes 2} - \frac{1}{6}\sum_{l=1}^6(-1)^l(\mathcal{M}_l^{\otimes 3})^{\otimes 2}\Bigr).
\end{align}
We then further apply the lifted witness $\mathcal{W}_2$ to two copies of the state $\rho^{ABC}$ and find that this indeed reduces to a nonlinear witness in the single-copy space:
\begin{align}\label{eq:wit2c}
    \tr(\mathcal{W}_2 \rho^{A_1B_1C_1}\otimes \rho^{A_2B_2C_2}) &=\, \frac{1}{2} \sum_{\substack{i,j,k=0\\0<i+j+k<3}}^1 \tr(\ketbra{ijk}{ijk}\rho^{ABC})^2-\frac{1}{12}\sum_{l=1}^6(-1)^l\tr(\mathcal{M}_l^{\otimes 3}\rho^{ABC})^2.
\end{align}

Now, we generalize this expression to $k$ copies and $N$ parties. To increase the number of copies we note that the generalized Hadamard map in the $Z$ basis acts as $\ket{0}^{\otimes k} \mapsto \ket{0}$ and $\ket{1}^{\otimes k} \mapsto \ket{1}$.
Then for three parties the witness lifted to the $k$-copy space reads
\begin{align}
    \mathcal{W}_k = \frac{1}{2}\Bigl(\sum_{\substack{i,j,k=0\\ 0<i+j+k<3}}^1 \ketbra{ijk}{ijk}^{\otimes k} - \frac{1}{3k} \sum_{l=1}^{3k}(-1)^l(\mathcal{M}_l^{\otimes 3})^{\otimes k}\Bigr),
\end{align}
which implies
\begin{align}
    \tr(\mathcal{W}_k\,\rho^{\otimes k})&=\,\frac{1}{2}\sum_{\substack{i,j,k=0\\ 0<i+j+k<3}}^1 \tr(\ketbra{ijk}{ijk}\rho)^{k} - \frac{1}{2}\frac{1}{3k} \sum_{l=1}^{3k}(-1)^l\tr(\mathcal{M}_l^{\otimes 3}\rho)^{k}.
\end{align}
Generalizing also to $N$ parties, we obtain
\begin{align}
    \mathcal{W}_k&=\,\frac{1}{2}\Bigl(\sum_{\substack{i_1,...,i_N=0\\ 0<i_1+...+i_N<N}}^1 \ketbra{i_1...i_N}{i_1...i_N}^{\otimes k} - \frac{1}{Nk} \sum_{l=1}^{Nk}(-1)^l(\mathcal{M}_l^{\otimes N})^{\otimes k}\Bigr),
\end{align}
or
\begin{align}
    \tr(\mathcal{W}_k\rho^{\otimes k})&=\,\frac{1}{2}\sum_{\substack{i_1,...,i_N=0\\ 0<i_1+...+i_N<N}}^1 \tr(\ketbra{i_1...i_N}{i_1...i_N}\rho)^{k} - \frac{1}{2}\frac{1}{Nk} \sum_{l=1}^{Nk}(-1)^l\tr(\mathcal{M}_l^{\otimes N}\rho)^{k},
\end{align}
in the single-copy space, where the respective measurements $\mathcal{M}_l$ have to be chosen according to Eq.~(\ref{eq:M}).  

Note that the lifting method works for any witness $\mathcal{W}$ and more
general projections than the Hadamard map, since one can always apply the inverse projection map to a given GME witness, lifting it to the $k$-copy space. The lifted witness can then be decomposed into a suitable basis and becomes therefore a nonlinear witness in the single-copy space:
\begin{align}
    \tr(\mathcal{W}\mathcal{F}(\rho^{\otimes k})) &=\, \tr(\mathcal{F}^*(\mathcal{W})\rho^{\otimes k})\,=\, \tr(\mathcal{W}_k \rho^{\otimes k})\,=\, \tr\Bigl(\bigl[\sum_{m_1,...,m_k}c_{m_1,..,m_k}\mathcal{G}_{m_1} \otimes \cdots \otimes \mathcal{G}_{m_k}\bigr]\rho^{\otimes k}\Bigr)\nonumber\\[1mm]
    &=\, \sum_{m_1,...,m_k}c_{m_1,...,m_k}\tr(\mathcal{G}_{m_1}\rho)\cdots \tr(\mathcal{G}_{m_k}\rho),
\end{align}
where $\mathcal{F}(\varrho_{ABC}) = [F_A\otimes F_B\otimes F_C] \varrho_{ABC} 
[F_A^\dagger \otimes F_B^\dagger \otimes F_C^\dagger]$ is some local projection from the $k$-copy space to the single-copy space and 
$\mathcal{F}^*(\mathcal{W})$ is its dual, which corresponds to the inverted projection.
For the particular choice of the GHZ-fidelity witness we can express the nonlinear witness into a sum of at most $1+kN$ terms, while for arbitrary witnesses more terms might be needed. 

\begin{figure}[t!]
    \centering
    \large{(a)}\includegraphics[width=0.42\textwidth]{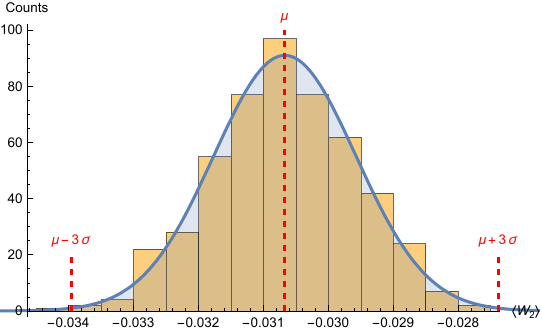}
    \ \ \ \ \large{(b)}\includegraphics[width=0.42\textwidth]{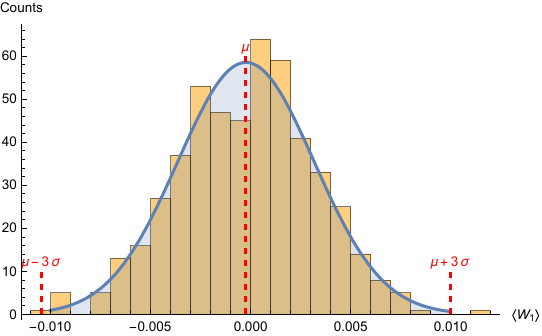}
    \caption{Example of the evaluation of the nonlinear witness for superactivation. For the model, we sampled $500$ times the measurement results of the $7$ measurements in Eq.~\eqref{eq:wit2c}
    for a $3$-qubit GHZ-state with white noise for the parameter 
    $p=3/7$. For representation, the data points are collected in bins and a Gaussian distribution with mean $\mu$ and standard deviation $\sigma$ obtained from the data points is displayed. 
    \textbf{(a)}  Results for the two-copy witness $\tr(\mathcal{W}_2 \rho^{\otimes 2})$ obtained from the measurement results of $Z$ and $\mathcal{M}_l$
    \textbf{(b)} Results for the one-copy witness $\tr(\mathcal{W}_{\mathrm{GHZ}} \rho)$ obtained from the measurement results of $Z$ and $\mathcal{M}_l$, showing  typically no
    violation (as expected), since the state is biseparable. 
    }
    \label{fig:sample}
\end{figure}

\subsection{Statistical model for a realistic experiment}

To demonstrate how the nonlinear witness can in principle be applied in an experiment, we consider the states used in Ref.~\cite{ChenEtAl2024}
\begin{align}
    \rho = p \ketbra{\mathrm{GHZ_+}}{\mathrm{GHZ_+}} + \frac{(1-p)}{8}\one_8.
\end{align}
This type of white noise is typically a good model for photonic experiments.

This state becomes biseparable for $p=3/7$, but the two-copy witness detects it as two-copy activatable. This leads to two conclusions: First, the state is indeed GME on the two-copy space. Second, the activated entanglement is compressible, and can be detected by applying locally a Hadamard-map and then checking the GHZ-fidelity witness, leading directly to a GME extraction and certification scheme. Indeed, in Ref.~\cite{ChenEtAl2024} the authors apply the Hadamard map for their distillation scheme and certify the existing entanglement using the GHZ fidelity. 

So, in order to certify that a given state is indeed useful for entanglement compression without implementing the whole distillation protocol, one can measure the nonlinear two-copy witness on the initial state. For the above scenario, one would need to measure the Pauli-measurement $Z^{\otimes 3}$ and the six measurements $\mathcal{M}_l^{\otimes 3}$ in Eq.~\eqref{eq:M}, leading to an expected value of $\tr(\mathcal{W}_2 \rho^{\otimes 2}) = -\frac{3}{98} \approx -0.031$. 

In order to estimate the statistical significance, we simulated $500$ times the measurement outcomes of the scenario where every measurement is measured $10,000$ times, and calculated the two-copy witness value from this. We find the distribution of the two-copy witness values pictured in Fig.~\ref{fig:sample} (a). The data points have an expectation value $\mu =-0.0307$ and a standard deviation of $\sigma = 0.0011$. This simple analysis implies that for a single run of the experiment one would obtain a mean value of $\mu =-0.031$ with a standard deviation of $0.025$.

Additionally, the measurement results of the above measurements can also be directly used to check the initial state for entanglement using the GHZ-fidelity witness (see, e.g., \cite{WeinbrennerPrasannan2024}). Using the same sample data we find the distribution of the single-copy witness pictured in Fig.~\ref{fig:sample} (b) with expectation value $\mu =0.000$ and a standard deviation of $\sigma = 0.004$, leading to a standard deviation of $0.08$ in one single run of the experiment. As expected, no violation can be observed since the state is biseparable. Of course, for proving biseparability more refined methods are needed, see, e.g., Ref.~\cite{SteinbergNguyenKleinmann2025} for related works.

%%%%%%%%%%%%%%%%%%%%%%%%%%%%%%%%%%%%%%%%%%%%%%%%%%%%%%%%%%%%%%%%%%%%%%%%%%%%%
\section{Explicit examples of incompressible entanglement}\label{app-sec:ICE}
%%%%%%%%%%%%%%%%%%%%%%%%%%%%%%%%%%%%%%%%%%%%%%%%%%%%%%%%%%%%%%%%%%%%%%%%%%%%%

We start by formally defining incompressible entanglement (ICE).
We say that a biseparable tripartite state $\varrho$ generates 
ICE (of level $k$) if the state $\varrho^{\otimes k}$ describing $k$ copies of it is GME, but any reduction to the single-copy space via 
local operations,
\begin{equation}
\tau_k = \frac{1}{\mathcal{N}}
[F_A\otimes F_B\otimes F_C] \varrho^{\otimes k} 
[F_A^\dagger \otimes F_B^\dagger \otimes F_C^\dagger],
\label{eq-app-icedef}
\end{equation}
is biseparable. Here, $F_X$ for $X=A,B,C$ are arbitrary local 
Kraus operators mapping a state from the $k$-copy space to a 
single-copy space and $\mathcal{N}$ denotes a normalization factor. 

Before continuing with our examples, let us  discuss some 
interpretational aspects. First, 
similar compressibility issues of quantum phenomena have been 
studied for measurements~\cite{BluhmRauberWolf2018, UolaKraftDesignolleEtal2021, LoulidiNechita2021}. 
Second, if one refers to bipartite entanglement, 
incompressibility effects can arise in 
several ways. As explained in the main text [see Eq.~(\ref{eq-icedistillation})], the question whether entanglement can be found
by projecting into subspaces of smaller dimensions is central for
the distillation problem~\cite{HorodeckiRudnickiZyczkowski2022}. Independently of that, one can consider
an entangled state in a $4\times 4$ system with a positive
partial transpose (PPT). Then, any projection down to a $2\times 2$
system is still PPT and hence separable. So, no entanglement can
be compressed to the two-level subspaces. In the present notion of 
Eq.~(\ref{eq-app-icedef}) it is therefore relevant that $\varrho$ is
biseparable, so the state $\varrho^{\otimes k}$ has a specific 
structure.  

This points towards an operational interpretation. Let us assume
that GME correlations on qubits are required for some task, but 
only the biseparable qubit state $\varrho$ is available. Then, the notion
of incompressibility decides about the usefulness of $\varrho$.
If one can prove that $\varrho^{\otimes k}$ is GME by projections
onto the single-copy state space, the state $\varrho$ may be useful. If, 
however, the state $\varrho$ generates ICE to any level $k$, it is
useless for a task requiring qubits. In this way, the notion of ICE may be seen in analogy to
the notion of bound (undistillable) entanglement.  

We now construct explicit examples of ICE states in the two-copy scenario. For this, we consider the state 
\begin{align}
\sigma = p \ketbra{\Psi^-}{\Psi^-} + \frac{(1-p)}{4} \one_4
\end{align}
on two qubits, where $\ket{\Psi^-}=\frac{1}{\sqrt{2}} (\ket{01} - \ket{10})$ is the singlet state. This state is invariant under any unitary transformation, i.e., $U\otimes U \ketbra{\Psi^-}{\Psi^-} U^\dagger\otimes U^\dagger = \ketbra{\Psi^-}{\Psi^-} $ for all unitaries $U$. We now consider two different families of states. The first family is given by
\begin{align}\label{eq-app:symmetricICE}
    \rho^{\rm (s)} = \frac{1}{3} \left( \sigma_{AB}\otimes \frac{\one_C}{2} + \frac{\one_A}{2}\otimes \sigma_{BC} + \sigma_{AC}\otimes \frac{\one_B}{2} \right),
\end{align}
where in each term two of the parties share the state $\sigma$. We consider a second family of states of the form
\begin{align}\label{eq-app:unsymmetricICE}
    \rho^{\rm (u)} = \frac{1}{2} \left( \sigma_{AB}\otimes \frac{\one_C}{2} + \frac{\one_A}{2}\otimes \sigma_{BC} \right),
\end{align}
where we only consider the two terms where A and B, and B and C share the state $\sigma$, respectively, leading to a slightly asymmetric configuration. Nevertheless, both of these families are invariant under any local unitary operation of the form $U\otimes U \otimes U$.

We now consider two copies of these states $\rho\otimes \rho$. Using the PPT-mixture approach, we find that the state $\rho^{\rm (s)}\otimes \rho^{\rm (s)}$ is two-copy GME for $p\geq 0.781$, while the state $\rho^{\rm (u)}\otimes \rho^{\rm (u)}$ is two-copy GME for $p\geq 0.708$. 

To show the ICE property, note that one does not necessarily need 
to project the state onto three qubits (that is, a $2\times 2 \times 2$ subspace).  If one can show that projecting the state down onto a strictly larger subspace than $2\times 2\times 2$ (e.g., a $3\times 4\times 4$ subspace) already renders the state separable for any possible projection, then the state will also be separable if one projects it further down to the $2\times 2\times 2$ space. 

We start by considering the projection $\tilde{P}$ for one party (say, Alice), which maps the four-dimensional two-qubit space to 
the two-dimensional one-qubit space. This can be seen as a projection 
onto a two-dimensional subspace, i.e.,
\begin{align}
\tilde{P} = \one_4 - \ketbra{\tilde{\psi}}{\tilde{\psi}}_{A_1A_2} - \ketbra{\tilde{\phi}}{\tilde{\phi}}_{A_1A_2},
\end{align}
where $\ket{\tilde{\psi}}_{A_1A_2}$ and $\ket{\tilde{\phi}}_{A_1A_2}$ span a two-dimensional subspace. Since any two-dimensional subspace on two qubits contains at least one product vector~\cite{SanperaTarrachVidal1998}, we can w.l.o.g. write the projection also as
\begin{align}
\tilde{P} = \one_4 - \ketbra{ab}{ab}_{A_1A_2} - \ketbra{\phi}{\phi}_{A_1A_2}.
\end{align}
Then, following the argument above, if the two-copy state becomes biseparable after applying only the partial projection $P = \one_4 - \ketbra{ab}{ab}_{A_1A_2}$, the application of the full projection $\tilde{P}$ will also render the state separable. We therefore focus
on $P$ and consider the projected state
\begin{align}
P_A \otimes \one_B \otimes \one_C \rho^{\otimes 2} P_A \otimes \one_B \otimes \one_C.
\end{align}
As the state $\rho$ is invariant under local unitaries $U\otimes U \otimes U$ and the entanglement properties of a state are not changed 
by local unitaries, the above state is separable if and only if 
\begin{align}
    (\one - \ketbra{00}{00})_A \otimes \one_B \otimes \one_C \rho^{\otimes 2} (\one - \ketbra{00}{00})_A \otimes \one_B \otimes \one_C
\end{align}
is separable, so it suffices to consider the projection $P = \one_4 - \ketbra{00}{00}$.

\begin{table}[t!]
    \centering
    \begin{tabular}{| l|| c | c | c |}
    \hline
      Criterion    &  $\varrho^{\rm (s)}$,  $P$ on $A$ &  $\varrho^{\rm (u)}$, $P$ on $A$ & $\varrho^{\rm (u)}$, $P$ on $B$ \\
      \hline
      Two-copy GME   & $p\geq 0.781$ & $p\geq 0.708$ &  $p\geq 0.708$ 
      \\
      \hline
      Projection PPT mixture   & $p\leq 0.800$ & $p\leq 0.738$ &  $p\leq 0.721$ 
      \\
      \hline
       Projection BS [Iter.]   & $p \leq 0.784$ & $p \leq 0.722$ & $p \leq 0.721$
      \\
      \hline
       Projection BS [Gilbert]   &  & $p \leq 0.720$ & $p \leq 0.721$
      \\
      \hline
    \end{tabular}
    \caption{Parameter ranges of the noise parameter $p$ for which the different states are two-copy GME and for which the states after projection become a PPT mixture or are detected by a biseparability
    (BS) algorithm. One finds in the three cases that (a) for $0.781 \leq p \leq 0.784$, (b) for $0.708 \leq p \leq 0.722$, and (c) for $0.708 \leq p \leq 0.721$, the respective states are ICE. In the first case, the Gilbert algorithm remained inconclusive for interesting values of $p$.}
    \label{tab:icetable}
\end{table}

Considering the state families from above, we find that $P\otimes \one\otimes\one (\rho^{\rm (s)})^{\otimes 2} P\otimes \one\otimes\one $ is a PPT mixture for $p\leq 0.800$ and GME for $p\geq 0.801$. For 
the second family we applied the projection either to party A 
or to party B. In the first case we find that 
$P\otimes \one\otimes\one (\rho^{\rm (u)})^{\otimes 2} P\otimes \one\otimes\one $ is a PPT mixture for $p\leq 0.738$ and GME 
for $p\geq 0.739$, while in the second case, the state 
$\one\otimes P\otimes\one (\rho^{(u)})^{\otimes 2} \one\otimes P\otimes \one $ is a PPT mixture for $p\leq 0.721$ and GME for $p\geq 0.722$.

For each of these states this leaves a considerable parameter regime where the full two-copy state is GME, but the projected state on a $3 \times 4 \times  4$-dimensional state space is a candidate for biseparability, 
since it is a PPT mixture. All that remains to be done is to prove
biseparability for some states in this regime and, luckily, for this task suitable tools exist \cite{Barreiro2010, Kampermann2012, Shang2018}. 
These tools deliver explicit biseparable decompositions, albeit with 
many terms, but require only a modest computational effort. We used
two different tools, an iterative algorithm \cite{Barreiro2010,Kampermann2012} as well as a version of 
Gilbert's algorithm \cite{Shang2018}. Both methods proved
the existence of ICE, see also Table~\ref{tab:icetable}.

\bibliography{references}

\end{document}